\documentclass[twocolumn]{aastex631}
\newcommand{\kms}{\ifmmode{\,\hbox{km\,s}^{-1}}\else {\rm\,km\,s$^{-1}$}\fi}

\shorttitle{Stream Heating}
\shortauthors{Carlberg \& Agler}
\begin{document}

\title{Sub-Halo Spreading of Thin Tidal Star Streams}

\author[0000-0002-7667-0081]{Raymond G. Carlberg}
\affiliation{Department of Astronomy \& Astrophysics,
University of Toronto,
Toronto, ON M5S 3H4, Canada} 
\email{raymond.carlberg@utoronto.ca}

\author[0000-0003-2429-8916]{Hayley Agler}
\affiliation{Department of Astronomy \& Astrophysics,
University of Toronto,
Toronto, ON M5S 3H4, Canada} 
\affiliation{Department of Physics \& Astronomy,
McMaster University,
Hamilton, Ontario,
L8S 4K1
Canada} 
\email{aglerh@mcmaster.ca}

\begin{abstract}
Dark matter sub-halos that pass a thin tidal star stream change the velocities of the stars near the point of closest encounter. Subsequent orbital evolution  increases the stream width and spreads the changes along the stream.   We measure the average widths and velocity dispersions of stream populations in three Milky Way-like CDM cosmology simulations containing star particle globular clusters and galactic disks of 0, 1 and 2 times a baseline model.     Power law fits to the velocity dispersion with stream longitude, $\phi$, for the overlaid streams in the 10-60 kpc range  find $\sigma\simeq$ 5-15 $\phi^{0.2-0.5} \kms$, with the coefficients depending on the sub-halo numbers, as well as the stream measurement details.  The velocity distributions summed along the entire length of the streams do not require the progenitor location. They  also rise with sub-halo numbers and  are significantly non-Gaussian with the ratios of the 6$\sigma$ to the 3$\sigma$ clipped velocity dispersions  being $\sim 1.5\pm 0.3$ and $\sim 2.5\pm 1$ for measurements within 1\degr\ and 5\degr\ of the streams, respectively.  Streams longer than 50\degr\ have an average radial velocity dispersion of 2.1 \kms\ with a disk  and 4.2 \kms\ without a disk. However, a few similar thin, low velocity dispersion streams are present in all three simulations. Statistically reliable conclusions require velocity data extending several degrees from the stream centerline for a set of streams.
\end{abstract}

\section{INTRODUCTION}

Thin tidal streams are powerful probes of the shape and the internal substructure of the dark matter which dominates the mass of the galactic halo  \citep{LBLB95,Morrison00,Ibata02,Johnston02}. The tidal field of the galaxy pulls stars away from the outskirts of a globular cluster, releasing  the stars to orbit freely in the galaxy.   Energy exchange between stars  within a cluster \citep{Chandrasekhar42, vonHoerner57, Henon61} boosts inner cluster stars to orbit out to the cluster outskirts, where tidal heating at pericenter passages \citep{Aguilar85,BT08,Meiron21},  to continue the mass loss from the clusters. The locations and velocities of the stars along the stream provides a snapshot of the potential along the stream's path from which the mean mass profile of the galactic halo and its large scale asymmetries can be derived \citep{Binney08}.

The sub-halo mass function in a galactic halo is a power law for cold dark matter \citep{Klypin99,Moore99},  $N(>M)\propto M^{-0.9}$  \citep{Springel08}. On the other hand, warm dark matter has a particle free-streaming cutoff scale \citep{BS83} which depletes the numbers of sub-halos below masses $\lesssim 10^9 M_\sun$ \citep{Benson13,Angulo13,Lovell14} with details depending on the dark matter particle properties.  The sizes and masses of the sub-halos set the scale size and velocity of the perturbation to a stream. The volume density of the substructures in the galactic halo determines the rate of encounters with the stream \citep{Ibata02,Johnston02,Carlberg12}.  Measurements of stream velocity dispersions then offer the prospect of a quantitative measurement of the sub-halo numbers. 

Finding  tidal streams immersed in the galactic background of unrelated stars requires well calibrated wide-field photometric data \citep{Irwin94,Gunn98,York00} and color-magnitude diagram filtering techniques \citep{Grillmair95,Odenkirchen01,Rockosi02}, leading to the discovery of a whole range of tidal streams from globular clusters and dwarf galaxies \citep{Ibata94,Belokurov06,GD1}.  The number of streams has now reached about one hundred and continues to grow \citep{Mateu22} with the Gaia Mission \citep{GaiaMission} and spectroscopic facilities providing proper motions and radial velocities \citep{DESI_MWS}, along with colors and spectra that can be used to distinguish stream stars from the field. Individual streams have been examined for density variations, gaps, along a stream, which  provides evidence that low mass $10^{6-8} M_\sun$  dark sub-halos are present \citep{Yoon11,Carlberg12} particularly in the exceptionally long and thin GD-1 stream \citep{GD1,Koposov10,CG13,Bonaca19}. Measurement of the properties of a population of streams  complements the detailed study of individual streams.

This paper analyzes simulations of sets of streams from star clusters in realistic halos to measure the evolution of the kinematics  of streams and characterize their current time properties. The elements of stream spreading have been studied in simulations of ever increasing complexity, from a sub-halo population in a spherical static halo \citep{Carlberg09}, a sub-halo population in an aspherical static potential \citep{Ngan15,Ngan16,Qian22,Ferrone23}  and here the sub-halos found in dynamically assembling halos which develop different sub-halo numbers. The simulation setup is described in \S\ref{sec_sims}. Section \ref{sec_props} discusses the kinematic properties of the streams. The basic dynamics of streams are reviewed in \S\ref{sec_dynamics}.  The spreading of the streams  with time is measured in \S\ref{sec_age}. Section~\ref{sec_subhalos} examines the role of sub-halos in the time evolution of individual streams. Predictions for the velocity distribution,  both over the visible length and with distance along a set of streams, are given in \S\ref{sec_obs}. Section~\ref{sec_discussion} summarizes the results and some implications for observational studies. 

\section{Simulation Setup \label{sec_sims}}

\begin{figure}
\begin{center}
\includegraphics[angle=0,scale=0.36,trim=30 10 100 80, clip=true]{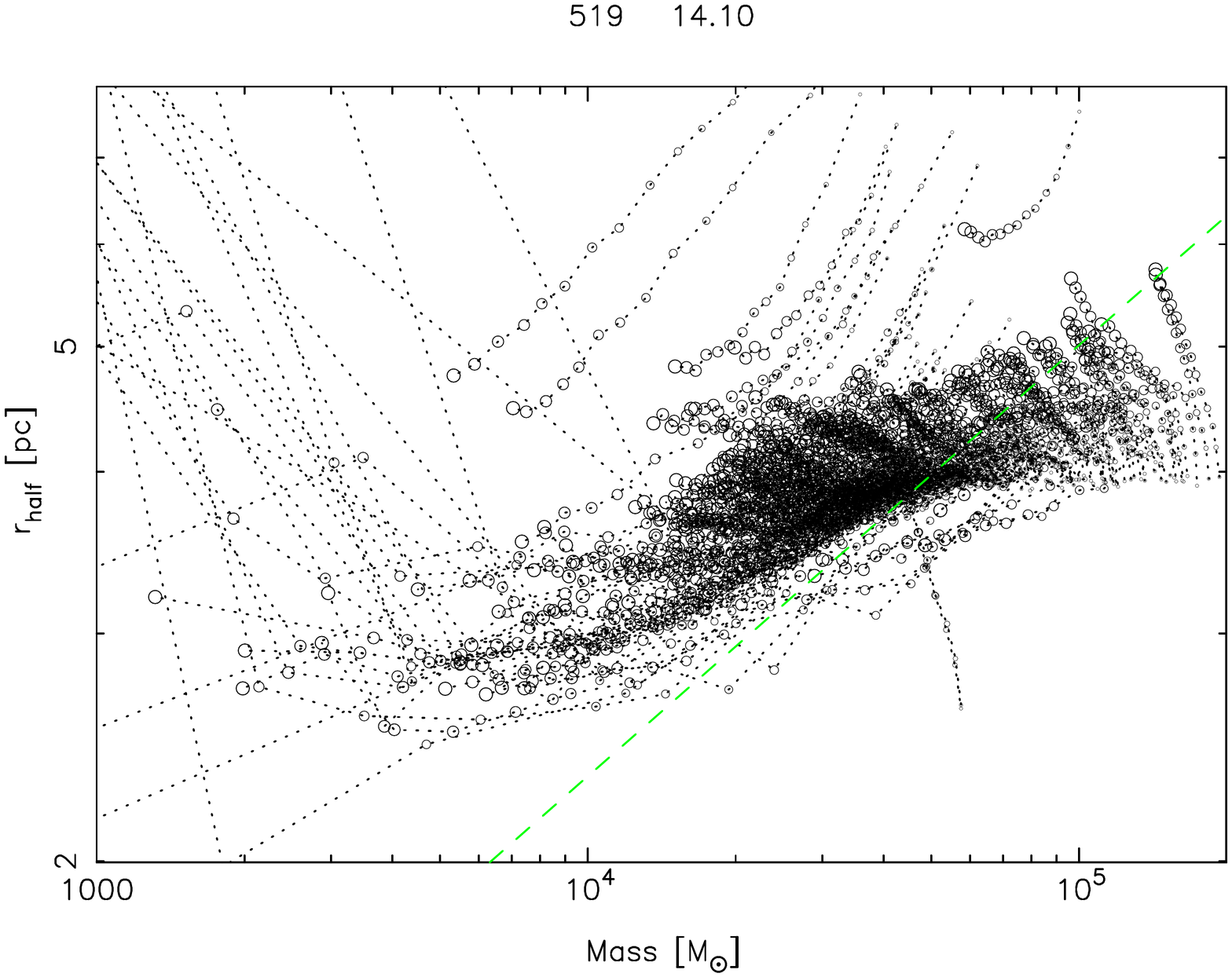}
\end{center}
\caption{
The mass radius relation of the clusters  with the circles along a cluster's path increasing in radius with time. The softening is 2 pc causing the lowest mass clusters to be artificially large. An average cluster loses about 60\% of its mass over the course of the simulation with some of the lowest mass clusters completely dissolving. The green line shows the $r_h=5 (M/10^5 M_\sun)^{1/3}$ constant density relation.
}
\label{fig_massrad}
\end{figure}

The simulations are started with the dark matter distribution used in \citet{CK22}, which was derived from the initial conditions of FIRE model m12i \citep{Hopkins14}.   The 70.514 million dark matter particles, each having a mass of $3.518\times 10^4 M_\sun$ are assigned a softening of 40 parsec. The early time particle distribution is evolved to  0.8 Gyr at which time the dark matter halos in the simulation were identified using the Amiga Halo Finder code \citep{AHF1,AHF2} as host dwarf galaxies for a population of high redshift, low metal abundance, globular clusters which are the progenitors of most thin streams \citep{Martin22}. Globular cluster-like star particle clusters are constructed using a W=7 King model \citep{King66} with masses randomly drawn from the range $4-20\times 10^4 M_\sun$ with an $N(M)\propto M^{-1}$ distribution. The clusters are placed in randomly oriented, rotating, exponential, disks in the inner 0.5 kpc of the dark halos, which have a typical radius of peak circular velocity of a few kpc. The total mass of star clusters in a halo is limited to $10^{-4}$ times the mass of the halo \citep{HHH14} for halos more than $5\times 10^8 M_\sun$.  The procedure generates 4.718 million star particles in 271 star clusters  placed in 133 dark halos. The outer radii of the clusters are started with the tidal radius at the radial positions of the clusters. The clusters quickly evolve into equilibrium as they orbit in their initial local sub-halo. The star particles have masses of $4 M_\sun$ with an assigned softening of 2 parsecs. The resulting star clusters have typical half mass radii, $r_h$, of 4 pc, Figure~\ref{fig_massrad}. The star particles are accurately followed from the half mass radius through the tidal zone out into the stream.   The softening allows reasonable time steps in the simulation and reduces the relaxation time into a regime that can be simulated accurately.  The code includes a cluster star relaxation process using  Monte Carlo velocity perturbation scheme \citep{Carlberg18}. The internal heating algorithm is calibrated to precise Nbody6 \citep{Aarseth99}  runs.  
The mass loss rate of a cluster depends on its mass, which increases with decreasing cluster mass,  Figure~\ref{fig_massrad}, and the tidal field along its orbit, which on the average is decreasing with time as accretion builds up the main halo. 

\begin{figure}
\begin{center}
\includegraphics[angle=0,scale=0.42,trim=20 30 60 60, clip=true]{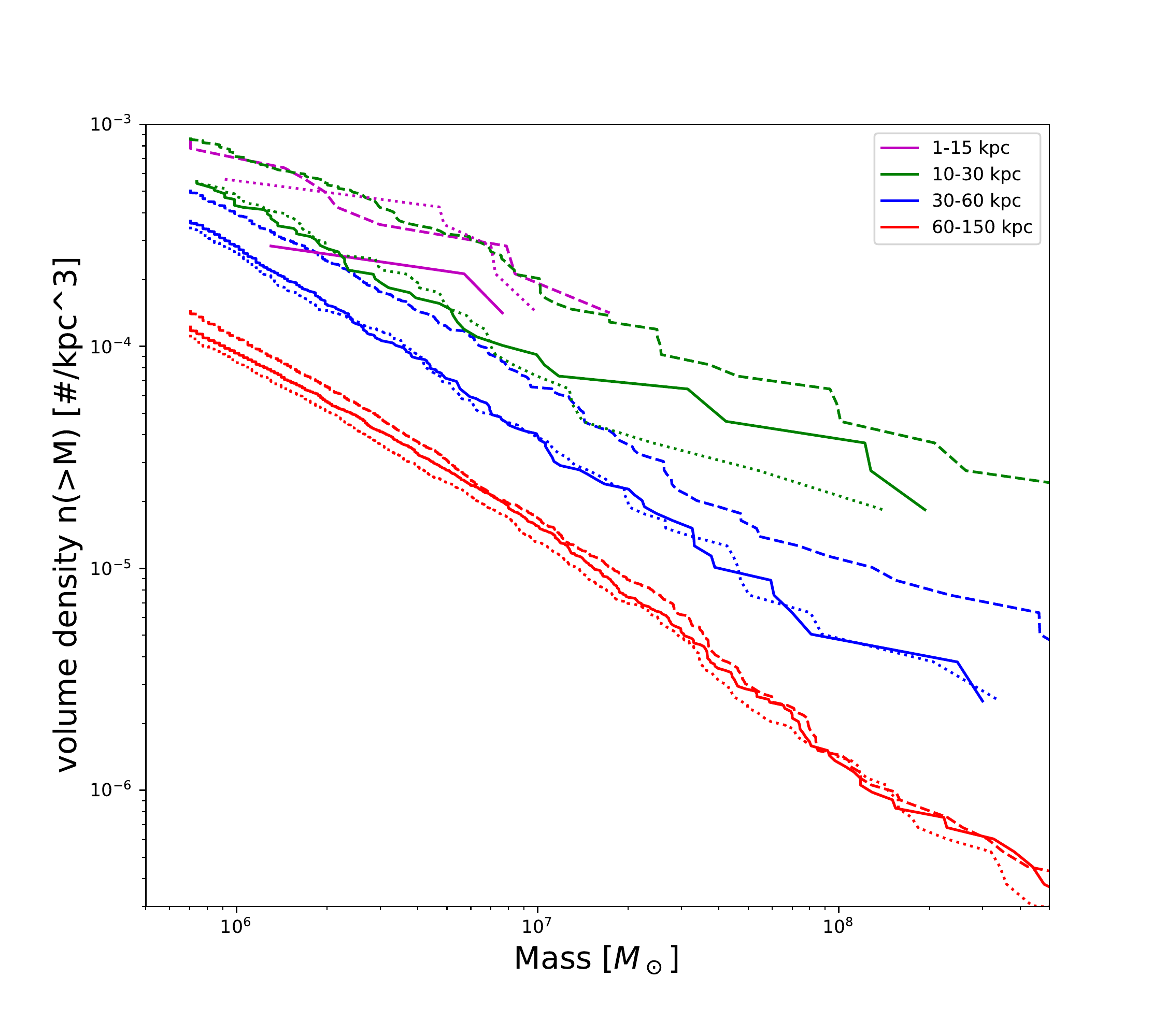}
\end{center}
\caption{
The cumulative volume density of sub-halos $N(>M)$ at the end time of the simulation with (solid lines) and without (dashed lines) a disk in four radial ranges.  The mass functions are close to $N(>M) \propto M^{-0.9}$. The solid lines are for the baseline simulation, dashed for the no-disk model, and dotted for the double mass model. 
}
\label{fig_halomf}
\end{figure}

\begin{figure}
\begin{center}
\includegraphics[angle=0,scale=0.48,trim=20 10 40 50, clip=true]{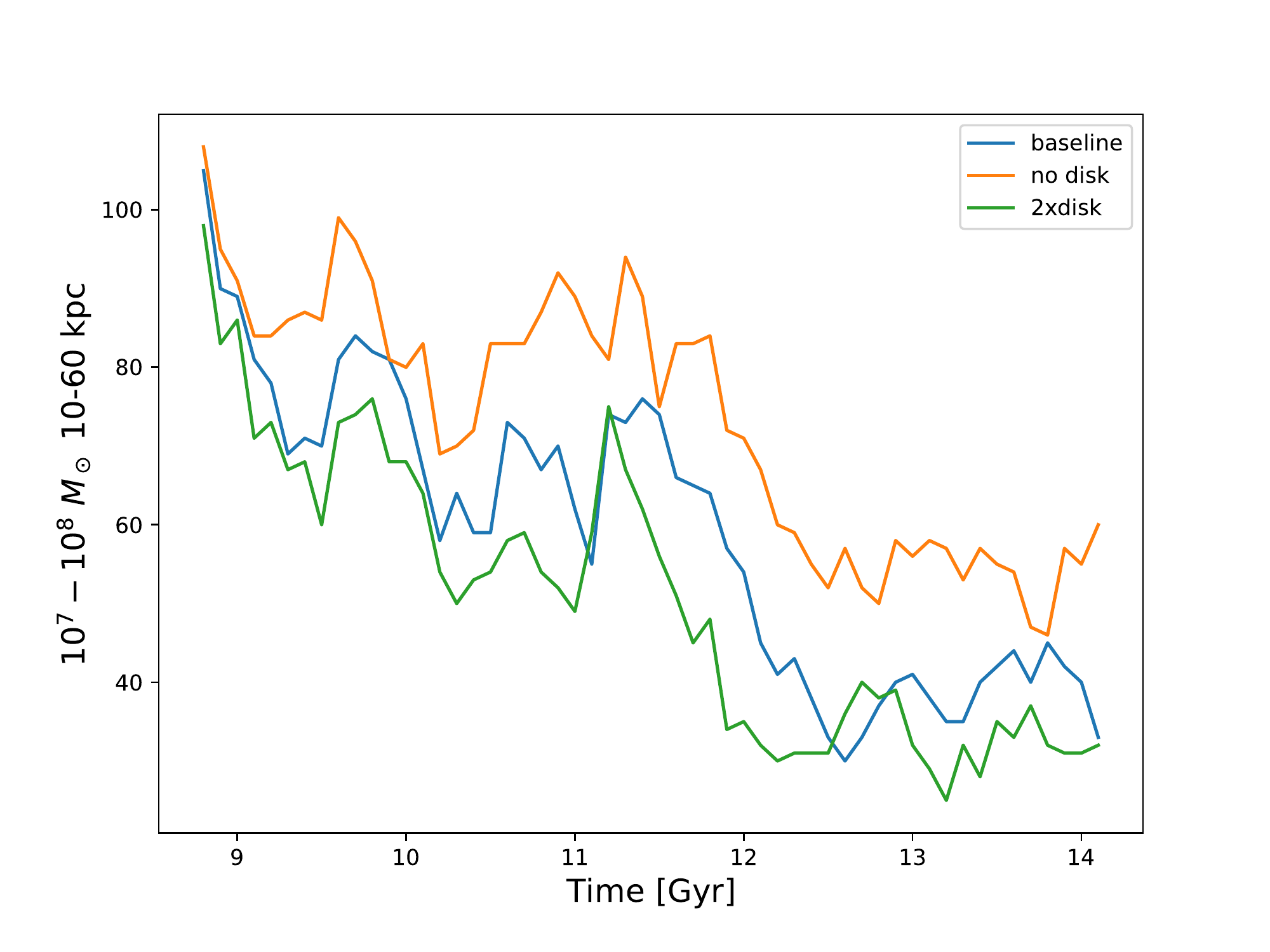}
\end{center}
\caption{The number of $10^{7-8} M_\sun$ halos between 10 and 60 kpc with time in the three simulations.
}
\label{fig_halont}
\end{figure}

\begin{figure}
\begin{center}
\includegraphics[angle=0,scale=0.45,trim=40 30 40 40, clip=true]{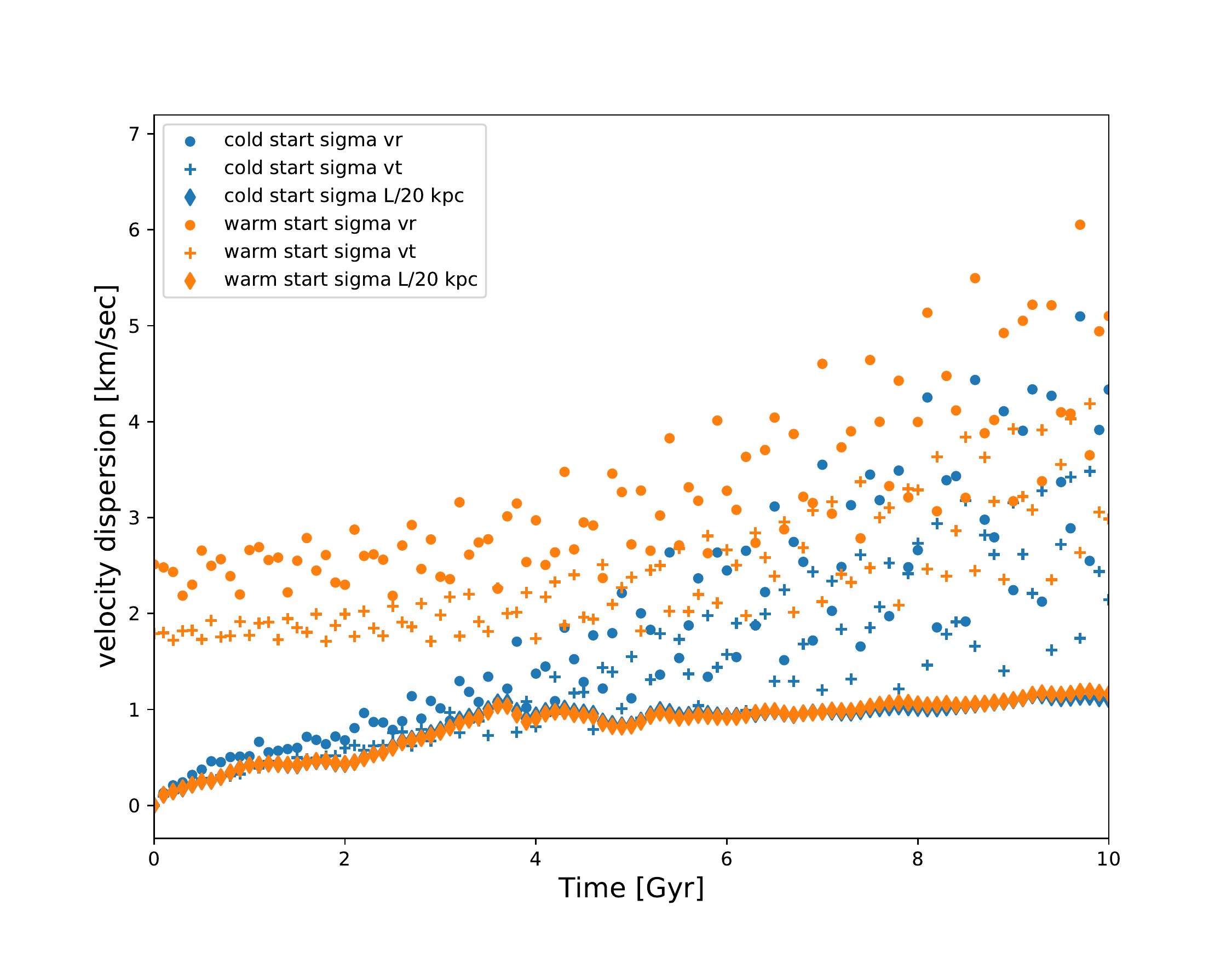}
\end{center}
\caption{
The characteristic velocity dispersions with time for a ring of 30,000 particles in the symmetrized main halo at the final moment. The radial and tangential velocity dispersions have a ratio approximately as expected from epicyclic motion, $\kappa/\Omega\simeq 1.4$ for a potential with a slowly varying circular velocity with radius, where $\kappa$ is the radial orbit frequency and $\Omega$ is the tangential orbital frequency.
}
\label{fig_dmheat}
\end{figure}

The largest  dark matter halo at 4.8 Gyr is augmented with a central bulge and a growing disk potential to create a  Milky Way-like main halo gravitational potential. A galactic disk potential depletes the sub-halos in the inner halo \citep{GKBullock17}. The disk also introduces an axis of symmetry which reduces the triaxiality of the inner halo. The dark mass of the main dark halo assembles very quickly, with 70\% of the mass of the main halo in place within 5 Gyr. The disk of the Milky Way grows at a much slower rate than the halo mass assembly,  which we approximate as linear in time, both in mass and in cylindrical radius \citep{FrankelRix19}.  The final mass of the main halo in the simulation is $5.7\times 10^{11} M_\sun$ within 100 kpc, slightly below the $6.1-6.9\times 10^{11} M_\sun$ of recent measurements \citep{Deason21,Shen22}. We start the disk growing at time 4.8 Gyr, shortly after a major merger and the beginning of the phase of relatively slow accretion growth of the main halo. A particle of mass $0.05\times 10^{10} M_\sun$ with a softening of 0.5 kpc is placed at the center of the halo. This particle is a lighter version of the nuclear bulge \citep{Launhardt02} which is not important to the dynamics at a few kpc but avoids small time steps near the center. A  \citet{MN75} disk of mass $3.4\times 10^{10} f(t) M_\sun$, with parameters $a= 2.0 f(t)$ kpc and $b=0.28$ kpc is centered on the bulge particle. The growth of both mass and disk scale length with time is 0 before 5 Gyr, after which it is $f(t) = (t-5)/9.1$ to a final time of 14.1.  The final disk is $3.4\times 10^{10} M_\sun$, slightly more than the $3.2\times 10^{10} M_\sun$ (combined thin and thick disks) of \citet{Zhou23}, but half the MW2014 model disk  \citep{galpy}.  The baseline and double disk models give 8.1 kpc circular velocities of 195 and  232 \kms, respectively. Three simulations, Table~\ref{tab_models}, are run, one with no disk and one with the \citet{galpy} disk (double our baseline),  to generate a range of sub-halo numbers in the inner halo and disk influence on streams. 

\begin{table}
\caption{Simulation Final Masses  \label{tab_models}}
\hspace{-35pt}
\begin{tabular}{| r |r|c|c|c|}
\tableline
Model & $a_{disk}$ & $M_{disk}$ & $M_{halo}$ & $M_{total}$ \\
\tableline
 disk & kpc &\multicolumn{3}{|c|}{$10^{10} M_\sun$  $\leq$100 kpc}  \\
\tableline
baseline & 2 & 3.4 & 53.5 & 56.9 \\
no disk  & 0 & 0 & 52.1 & 52.1 \\
2$\times$ disk & 3 & 6.8 & 54.6 &  61.5 \\
\tableline
\end{tabular}
\end{table}

Figure~\ref{fig_halomf} shows the mass functions of the sub-halos in the main  halo at the end of the simulation for several radial bins. The mass distributions are very near a power-law of $N(>M)\propto M^{-0.9}$ \citep{Springel08}. The solid lines are for the baseline model with a disk. The dashed line is for a model started from the initial dark matter distribution, but with no added disk. The dotted lines are for a model with a disk twice the mass of the baseline model with the disk scale length boosted from 2 to 3 kpc to maintain a similar surface density in the two runs.  Figure~\ref{fig_halont}  shows the numbers of inner halo sub-halos in the $10^{7-8} M_\sun$ mass  interval and between 10 and 60 kpc.  The simulations with disk deplete the numbers of inner sub-halos relative to a disk-less simulation \citep{GKBullock17}. Accretion brings in new sub-halos at a declining  rate of addition in a noisy process.

\begin{figure}
\begin{center}
\includegraphics[angle=0,scale=0.32,trim=50 70 70 70, clip=true]{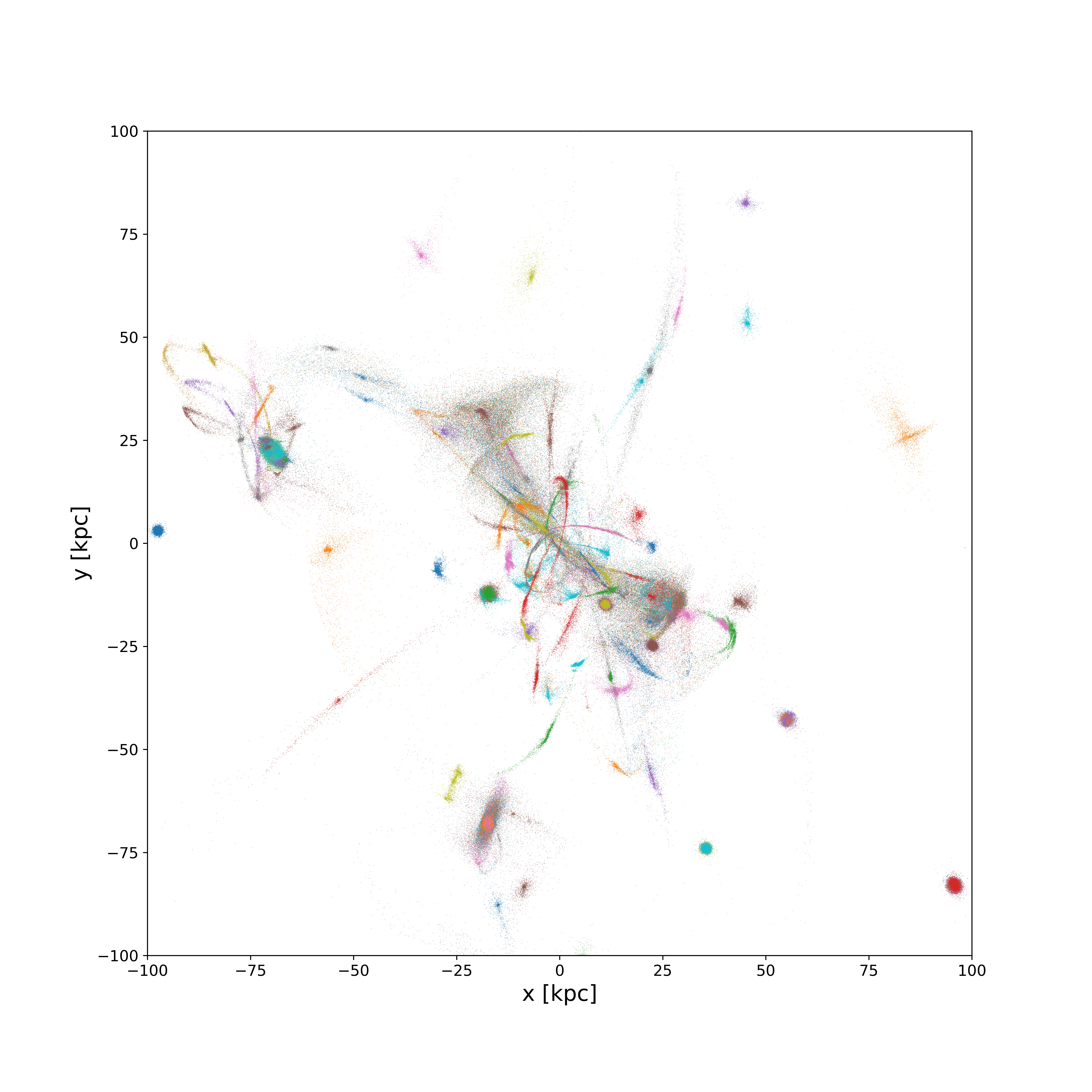}
\end{center}
\caption{
An x-y projection of the stars released from the clusters in the last 4 Gyr. 
}
\label{fig_xy}
\end{figure}

\begin{figure*}
\begin{center}
\includegraphics[angle=0,scale=0.45,trim=180 70 130 80, clip=true]{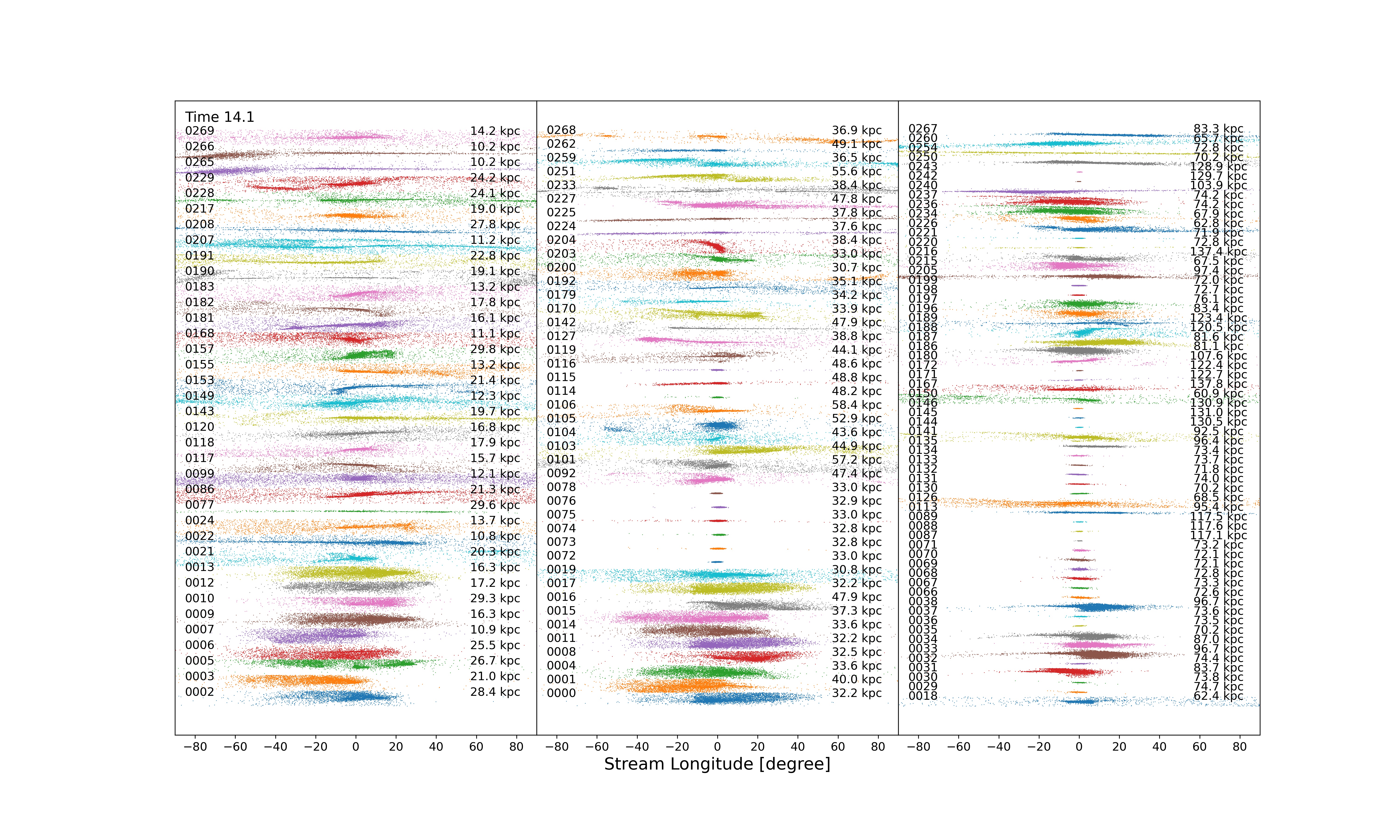}
\end{center}
\caption{All the streams within a distance of 10-150 kpc rotated into the frame of their progenitor clusters' great circle orbital plane.  The star particles within  a strip of $\pm 90 \degr$ wide and $\pm 30\degr$ wide  along  the local equator are plotted. The cluster numbers are indicated on the left and the current distance from the center is on the right. Many of the very short streams at large distances are orbiting within a sub-halo.
}
\label{fig_rotstreams}
\end{figure*}

The graininess of the n-body simulation causes the dark matter particle population to slightly increase the velocity dispersion of stream star particles. The  process is a random walk so the velocities increase approximately in proportion to the square root of time \citep{Chandrasekhar42}. \citet{Penarrubia18,Penarrubia19} has shown that the tidal field of the nearby particles dominates the increase in velocity dispersion of small coherent structures such as star clusters and streams. The effects are coherent over the inter-particle spacing of the dark matter particles, which is in the range of 0.1-1 kpc for the radial range of 3-100 kpc in our simulations. Streams respond to particle passages with coherent velocity changes over these length scales, which over several orbital times develop into density variations on similar scales in the streams \citep{BanikBovy21}.  We undertake a special purpose simulation  to measure the particle noise driven velocity rise for our simulation.  The particles from the final time main halo in the baseline simulation are placed at their radial locations but randomly rotated on a sphere to symmetrize and smooth the mass distribution. A ring of 30,000 star particles on circular orbits is started at 20 kpc, one simulation with a ring  at the local circular velocity and no random velocities,  another with Gaussian perturbations in radial velocity. The results are shown in  Figure~\ref{fig_dmheat}. The simulations show that the angular momentum is conserved to about 1.1\% over 10 Gyr, and the radial and tangential velocities rise from zero to about 3 \kms\  over 10 Gyr. Increasing the softening of the dark matter particles to as much as 1 kpc provides only a modest reduction in the particle heating, as expected from the logarithmic behavior of two-body scattering. The particle noise induced velocity changes are below the perturbations from sub-halos for these simulations but will add noise to streams as cold as 1 \kms.

The Gadget-4 code \citep{Gadget4} evolves the simulation. The code is augmented to include the slow heating of star particles in clusters with the algorithm of \citet{Carlberg18}. The gravitational field of the growing galactic disk is added to the gravitational force.  The small sizes of the star clusters require small time steps, typically 5-10 thousand years, so the simulation requires more than a million time steps to complete. Time is reported in computational Gyr time units, where 1 unit is 1.022 true Gyr. 

\begin{figure}
\begin{center}
\includegraphics[angle=0,scale=0.47,trim=10 20 40 50, clip=true]{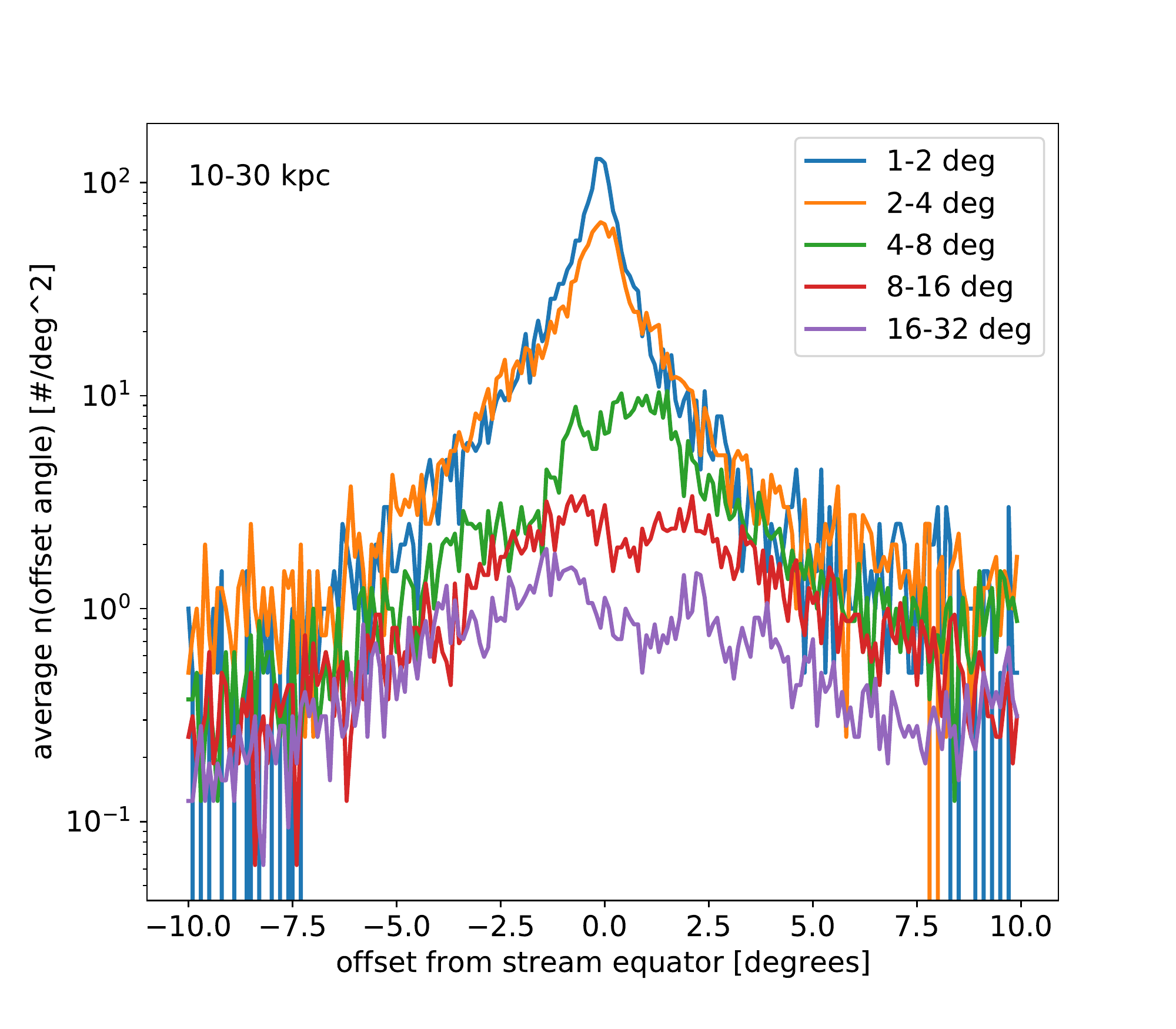}
\includegraphics[angle=0,scale=0.47,trim=10 20 40 50, clip=true]{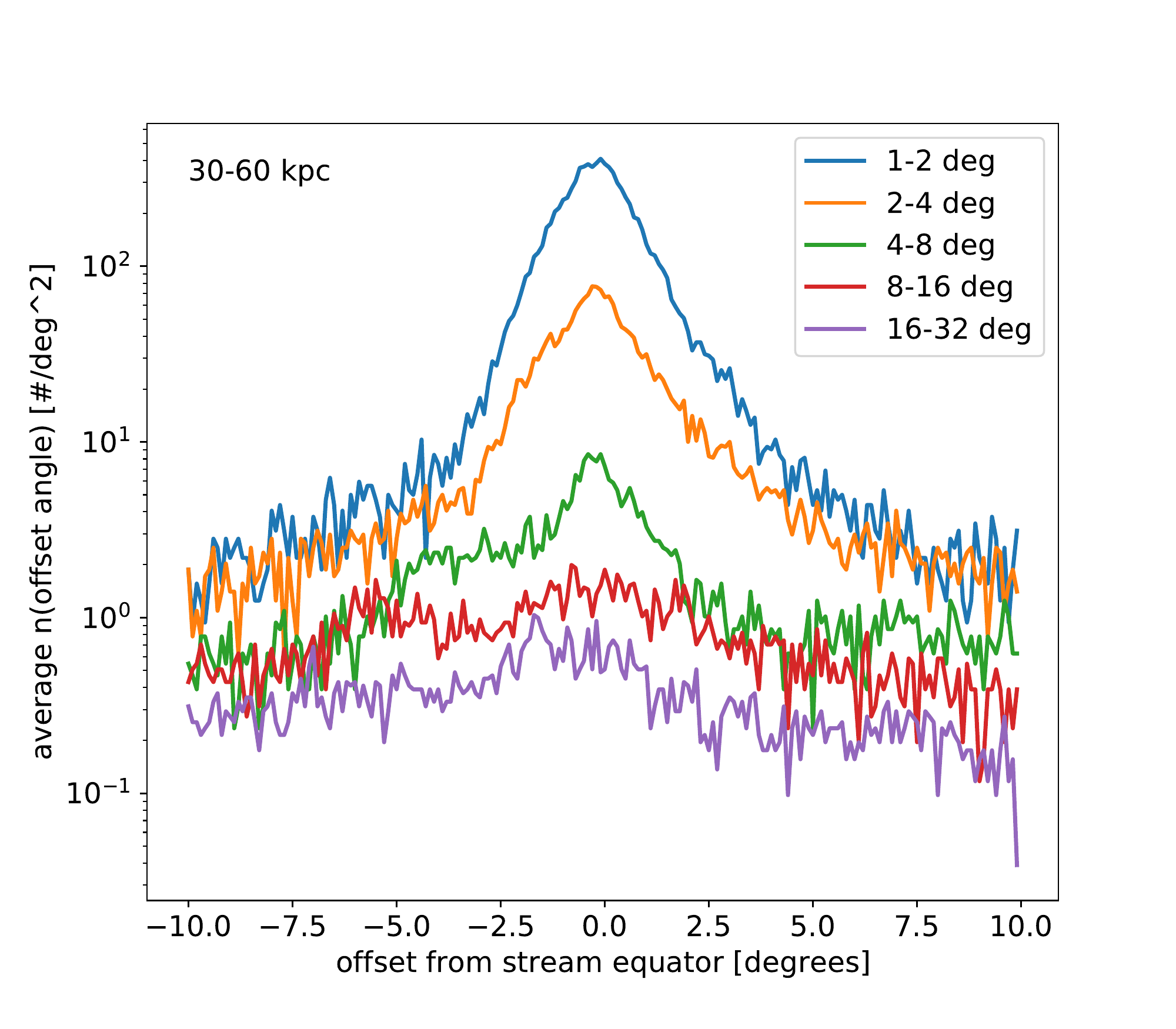}
\end{center}
\caption{
The average distribution of sky densities perpendicular to the streams' axes, for a range of distances along the streams. The streams have progenitors in the 10-30 kpc (top) range and 30-60 kpc (bottom) with $\left|L\right|>1000$ kpc-\kms.
}
\label{fig_widthdist}
\end{figure}

\section{The Streams on the Sky\label{sec_props}}

The x-y projection of the stars released from the clusters in the last 4 Gyr are shown in Figure~\ref{fig_xy}. The streams in the inner 60 kpc have orbital shapes ranging from nearly circular to nearly radial, with a tendency for more radial streams at larger radii. The set of streams at (-70, 22, -11) on the left middle of the plot are in a sub-halo of $1.45\times 10^{10} M_\sun$ which passed within 15 kpc and is now at 74 kpc, continuing to move away at 150 \kms. We focus much of our analysis on the region closer than 60 kpc, partly to avoid this structure and its clusters which orbit within its sub-halo and are not yet centered on the main halo. Known non-dwarf galaxy streams have median distances within 50 kpc \citep{Mateu22}.

The streams with progenitor clusters currently between 10 and 150 kpc are plotted using local stream coordinates in Figure~\ref{fig_rotstreams}. The inner limit at 10 kpc is to exclude streams which encounter the galactic disk, which completely disperses about half the streams. Each stream of Figure~\ref{fig_rotstreams}  has been rotated into its instantaneous great-circle frame which uses the velocity of the cluster  (a reference particle at the center of each cluster in the simulation) to define the equator of the orbital plane perpendicular to the cluster's angular momentum, and the current cluster position defines the  zero of the stream longitude.   Clusters IDs are shown on the left and their current distance on the right of Figure~\ref{fig_rotstreams}. The great-circle approximation  transforms the somewhat jumbled distribution of streams of Figure~\ref{fig_xy} into orderly, nearly linear, distributions. The inner streams have greater angular lengths and are more widely dispersed than those at large distances from the center, however there is a range of stream lengths and widths at all radii. In the simulation setup, clusters were put into host halos in order of halo mass, so there are sets of clusters that have remained close in radius, such as 72-78 in the middle panel. The streams generally have a thin core, surrounded by a wide component \citep{Carlberg18,Malhan19}. 

The average sky density of star particles (each 4 $M_\sun$ here) perpendicular to the stream axis is shown in Figure~\ref{fig_widthdist} for the 10-30 and 30-60 kpc distance ranges.   Near the progenitor cluster the streams have an average half-width at half-maximum of 0.5\degr.  The thin cores of the streams have a non-Gaussian tail around them extending beyond 10\degr\ in width. The asymmetries of Figure~\ref{fig_widthdist} are likely due to the small numbers of streams in the average, with the streams at 30-60 kpc showing a tilt in the opposite direction. On the average, streams become significantly wider with distance along the stream, which lowers the peak surface density of stars in the stream. The average surface density declines approximately a factor of 5 from 1\degr\ to 32\degr\ along the stream length. For an individual stream the density along the stream varies with  orbital phase with streams being stretched in longitude but narrowed in latitude at their orbital pericenter.  The density along each stream varies with orbital phase, reaching a minimum at pericenter, but being at a smaller orbital radius is often observationally easier to find. On the average, at an angular distance down the streams of 10\degr\ the mean stream surface density is about 10\% of the initial stream and the distribution becomes much broader, nevertheless most streams do have a thin core over some fraction of their length. 

\section{Stream Dynamics\label{sec_dynamics}}
 
This section uses the basic dynamics of streams to estimate the expected effects of the measured sub-halos on the streams. Tides pull stars away from the cluster into orbits that are offset outward and inward from the cluster by  approximately the Jacobi radius (\citet{BT08} Equation 8.14),
\begin{equation}
 r_J =  r\left({m_c}\over{2M(r)}\right)^{1/3} =  r\left({{Gm_c}\over{2v_c^2 r}}\right)^{1/3}.
 \label{eqn_tidal}
 \end{equation}
where $m_c$ is the mass of the cluster, $r$ is its galactic radius, $M(r)$ is the total galactic mass inside $r$, and $v_c$ is the circular velocity at $r$.  The time varying gravitational field of the simulations pulls star particles away from the clusters with a range of angular momenta. For this calculation, we  make the approximation that stars are added to the leading and trailing streams with the same velocity as the cluster but offset radially inward and outward by the tidal radius.  The stars pulled inward have less angular momentum than the cluster and will orbit at a higher angular rate, developing the leading tidal stream, and vice-versa for the trailing stream.  The stars that escape at zero relative angular velocity move away from the progenitor cluster at a velocity $\Delta v_s\simeq\Omega r_J$, where $\Omega=v_c/r$,
\begin{equation}
 \Delta v \simeq      \left({{Gm_cv_c}\over {2r}}\right)^{1/3} .
 \label{eq_dvtaill}
 \end{equation}
Stars leaving the cluster have a spread of velocities, $\sigma_r$, comparable to $\Delta v$, because the tidally induced velocities are proportional to the tidal acceleration, which varies linearly with distance from a cluster's center. Consequently,  the leading edge of the stream will move down the stream at $\sim 2\Delta v$. For typical values $m_c= 10^5 M_\sun$, $v_c= 200\kms$ and an orbital pericenter of $r= 10\, {\rm kpc}$, $2\Delta v= 3.2\kms$, which over 10 Gyr develops into a tidal tail of 16 kpc, for a total length of leading and trailing of 32 kpc.  The stream width is $\simeq \sigma_r/\kappa\simeq r_J$, where $\kappa$ is the radial oscillation frequency, equal to the epicyclic frequency for small eccentricity orbits.  Most stars are stripped from the outskirts of a star cluster at its orbital pericenter.  Streams are thinnest and most visible as they pass near their orbital pericenter.  The angular length of a stream should then be about $32/10\simeq 3.2$ radians or 180\degr, that is,  streams should typically wrap halfway around the sky. The four longest streams attributed to star clusters are 100-140\degr\ long (GD-1, M68, Murrumbidgee, NGC3201) with Orphan at 230\degr\ long possibly having a dwarf galaxy progenitor \citep{Mateu22}. The decline of surface density and increasing width of streams along their length seen in observed streams \citep{Bernard16,Bonaca20}  is also seen in the simulations. That is, the simulated streams generally have members around their entire orbit, however the stars most distant from the progenitor will be observationally difficult to find as their surface density declines relative to unrelated field stars.

\begin{figure}
\begin{center}
\includegraphics[angle=0,scale=0.44,trim=10 70 40 80, clip=true]{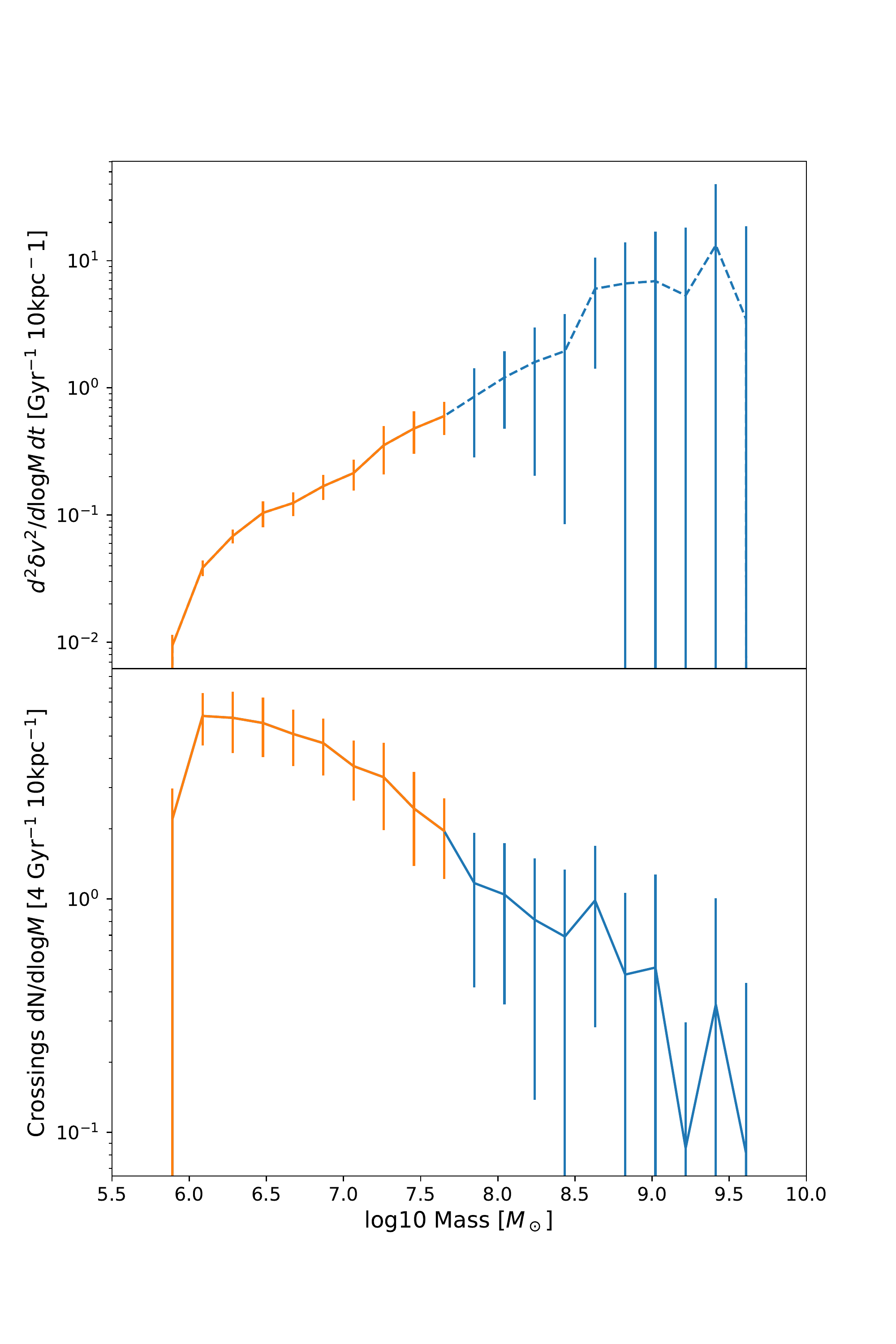}
\end{center}
\caption{
The top panel shows the differential velocity heating rate  in mass bins, evaluated with the impact approximation in the 10-60 kpc range, averaged over the last 4 Gyr. The bottom panel shows the expected number of sub-halo stream crossings for a 10 kpc length of stream over the last 4 Gyr. The orange colored lower mass part has more than two sub-halo encounters over the 4 Gyr interval. Higher sub-halo mass encounters are relatively rare, leading to coherent stream deflections.
}
\label{fig_heatnum}
\end{figure}

Sub-halos crossing the stream at velocity $v$ change the velocities of stream stars. An approximate analysis of the resulting stream heating helps to understand the process. The time for lower mass sub-halos to cross a stream is so short that the impact approximation is  satisfied, $\delta v \simeq 2GM_h/(bv)$, which we approximate as $2GM_h/(r_m v)$, where $r_m$ is the radius of the peak of the circular velocity curve for a sub-halo. The velocity change  for  the more massive sub-halos,  $\gtrsim 10^7 M_\sun$, which have virial radius over 4 kpc. will have increased dependence on the orbital details.  Fitting to the sub-halo peak circular velocity radii in the simulation we find a sub-halo changes the velocities of the nearest stream stars by,
\begin{equation}
\delta v = 1.3  \left( {M_h\over{10^8 M_\sun}} \right)^{0.49} \left({{300 }\over{v }}\right) \kms,
\label{eq_deltav}
\end{equation}
 with 10\% error on the velocity scale and 6\% on the power coefficient. The power law fit to the radius where the halo circular velocity peaks is  $r_m=(1.4\pm0.2)(M_h/10^8 M_\sun)^{0.48}$ kpc.  

The rate at which the velocity dispersion rise is  then  (see \citet{Carlberg13} for details),
\begin{equation}
{{d^2\langle{\delta v^2}\rangle}\over{dM_h\,dt}} =  \int\int {{4G^2M_h^2}\over{b^2 v^2}}{{dn(M_h,r)}\over{dM_h}}v f(v)\pi\, db\,dv,
\label{eq_heating}
\end{equation}
where $f(v)$ is the velocity distribution function,  $\int f(v)dv=1$, in the frame of stream and $n(M_h,r)$ is the radial density distribution of sub-halos of mass $M_h$. The virial radius is 3-6 times $r_m$, so we integrate to an impact parameter $b=3 r_m$, which is close to the virial radius, 
\begin{equation}
{{d^2\langle{\delta v^2}\rangle}\over{dM_h\,dt}} \simeq  20\pi{{G^2M_h^2}\over{r_m v}}n(M_h,r),
\label{eq_heating2}
\end{equation}
Combining the estimate of $\delta v$ of Equation~\ref{eq_deltav} with the measured halo mass function of approximately $n(M)\propto M^{-1.9}$  leads to a heating rate  proportional to $M^{0.14}$. 

The heavier sub-halos have a slight margin in heating  stream stars; however their effect diminishes as their encounters with a stream become less likely to occur. When there are less than two sub-halo stream crossings in some time interval a single sub-halo stream crossing leads to coherent stream changes, not heating. The 10-60 kpc radial distance range  dominates stream heating, with a total heating up to and including  the two-crossing mass of $10^{7.6} M_\sun$  (for 4 Gyr) of   $\sqrt{\langle\delta v^2\rangle}= 0.3 (t/{\rm Gyr})^{1/2}$ \kms.  Figure~\ref{fig_heatnum} averages heating and stream crossing rates over the last 4 Gyr at intervals of 0.2 Gyr. The plotted error bars give the RMS spread in the heating rate. About half  of the heating comes from sub-halos between  $10^7$ and $5\times10^7 M_\sun$.   We note that the simulated lower mass sub-halos may be artificially large and too easily disrupted \citep{Errani21}, so the heating from lower mass halos is likely somewhat underestimated.  

The calculations summarized in Figure~\ref{fig_heatnum} give a total heating of $\sim 1 \kms$ over 10 Gyr to the two-crossing sub-halo mass of $M\sim 10^{7.6} M_\sun$, beyond which sub-halo stream crossings are rare and the random quadrature sum is not valid.  Hence,  sub-halo heating alone cannot by itself explain all of stream spreading. However, in a triaxial or more complex potential, which is generally the case, the sub-halo perturbations  large scale potential asymmetries along the streams add yet more spreading \citep{Carlberg15,Ngan16}.

A  $10^{7.5} M_\sun$ sub-halo has $r_{virial}\sim 5$ kpc, so the dominant sub-halos will lead to coherent velocity changes over 10 kpc sections of a stream. The velocity perturbations will lead to orbital changes which will add to these velocities and mix the stars along the length a stream. Exactly how the velocity perturbations play out to measurable changes in the streams will be examined in the next sections. 

\begin{figure*}
\begin{center}
\includegraphics[angle=0,scale=0.43,trim=110 60 120 50, clip=true]{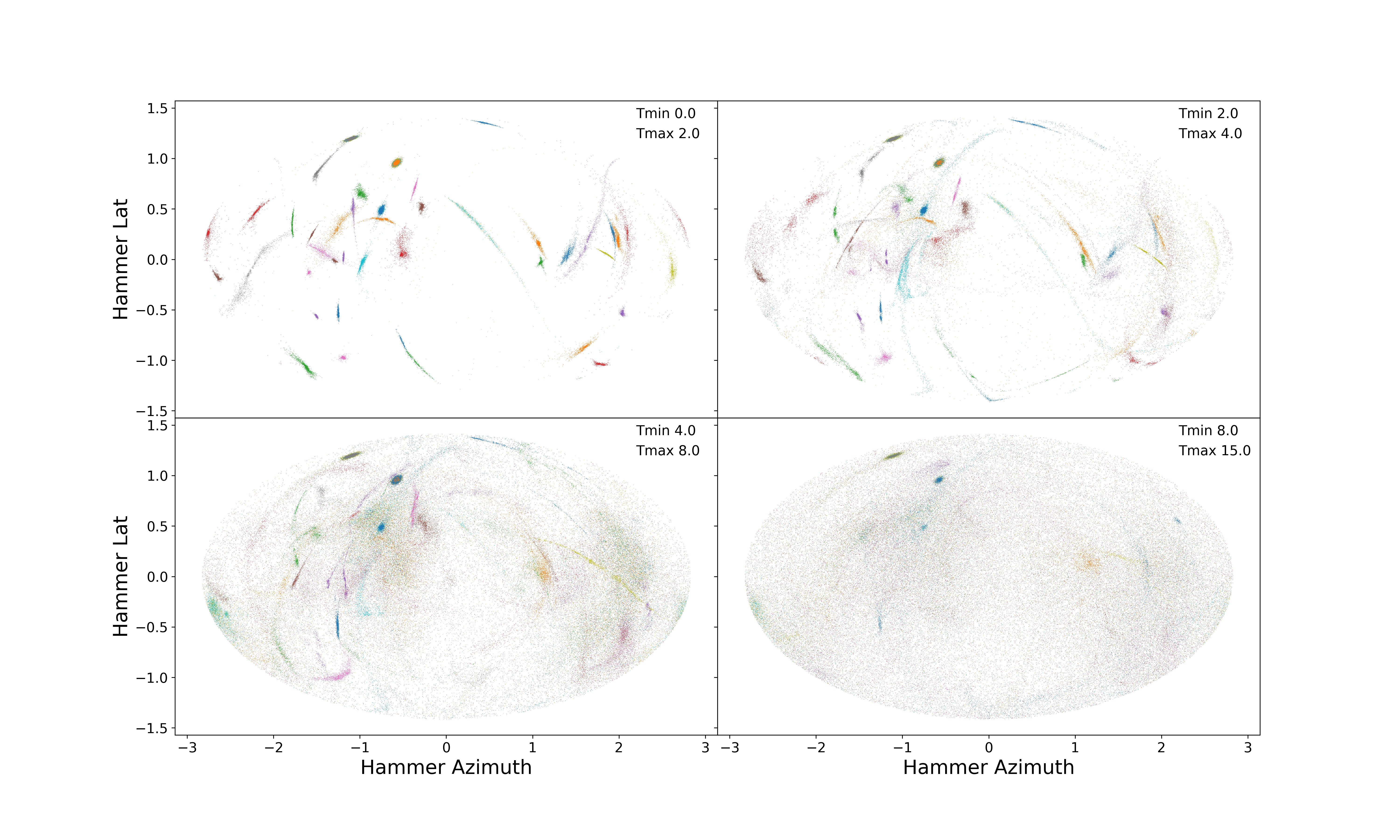} 
\end{center}
\caption{The  streams plotted on the sky from the galactic center point of view. The panels show particles released in the last 2 Gyr, 2-4 Gyr, 4-8 Gyr and 8 and more Gyr for clusters between 10 and 60 kpc.
}
\label{fig_hammer}
\end{figure*}

\section{Stream Spreading with Age \label{sec_age}}

The time at which a star leaves the cluster is measured in a simulation as the first time at which it is beyond some minimum distance from the cluster center.  We use 0.2 kpc as the minimum distance, which is typically twice the largest tidal radius for clusters in the main halo.  The age of the star in the tidal stream is the difference between the current simulation time and the time at which the star crossed the minimum distance radius.  Figure~\ref{fig_hammer} plots the sky positions of the stream stars in four age ranges from clusters that are in the radial range of 10 to 60 kpc with a minimum angular momentum of $\left|L\right|>1000$ kpc-\kms . The selected radial range reduces the numbers of streams that pass through the disk and those orbiting in infalling sub-halos. The stream age range of the stars is indicated in the legend for each panel, with the youngest stars in the upper left and the oldest in the lower right.  

The length on the sky of the stream stars increases with their age in streams, as expected for stars that are being pulled into the leading and trailing tidal tails. The older stars are in longer stream segments, but with a considerable overlap with the younger stars, a consequence of the spread in angular momentum of the stream stars. After about 4 Gyr the stars dramatically  spread out from the streams, with streams largely blurred out on the sky after 8 Gyr. The orbital perturbations are larger than the approximate calculation of Equation~\ref{eq_heating2} and  larger than seen in a static symmetric halo or aspherical potentials with sub-halos \citep{Ngan16}, indicating that it is the combination of sub-halo heating and orbit dispersion in the aspherical and time variable galactic halo potential.

\begin{figure}
\begin{center}
\includegraphics[angle=0,scale=0.22,trim=70 260 100 250, clip=true]{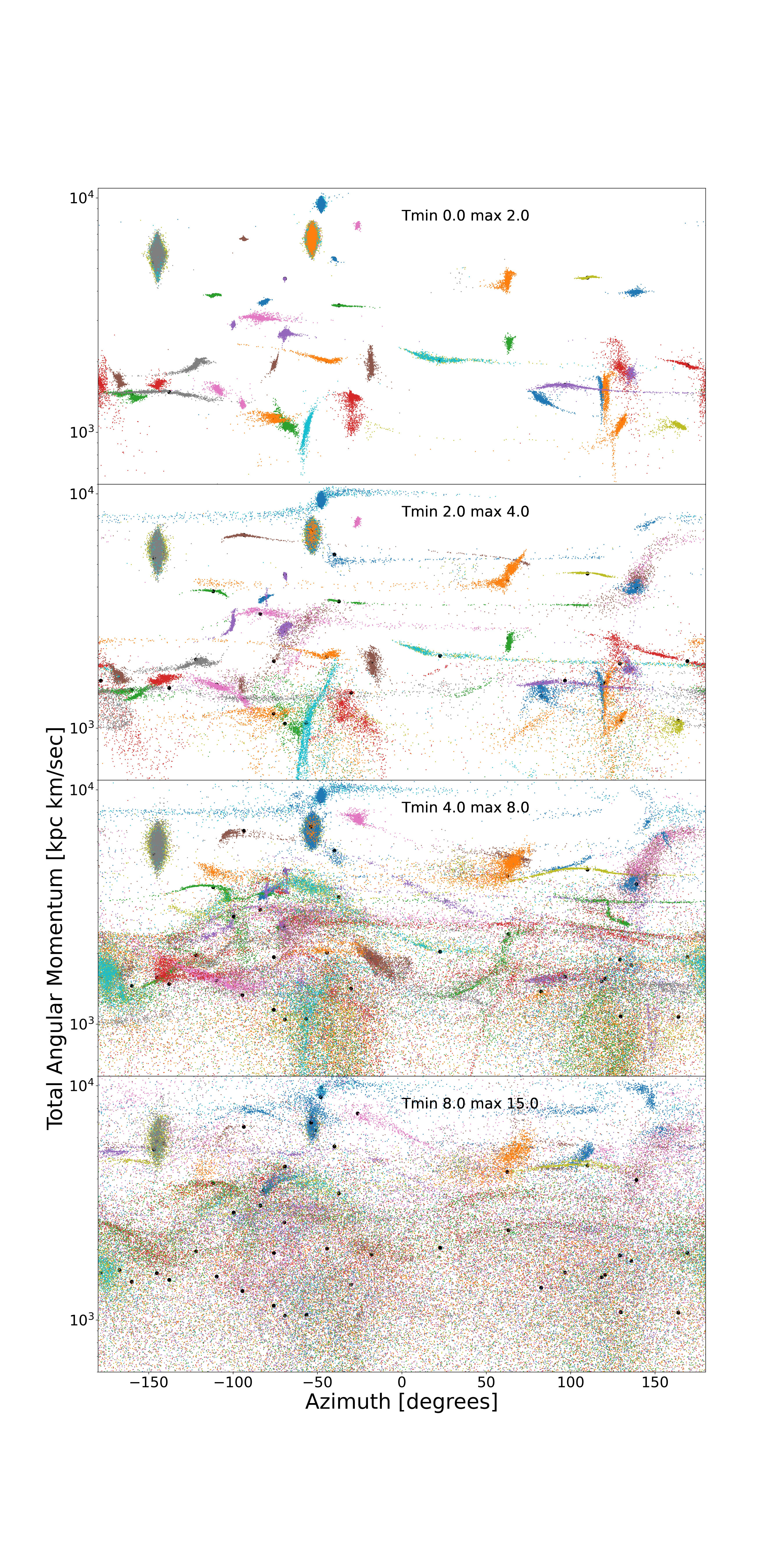}
\end{center}
\caption{The  angular momenta, $L$, of the stars are plotted against the longitude on the sky, for clusters with $L>1000$ kpc-\kms\ and  in the radial range 10-60 kpc.  The top panel is for stars that left their clusters  within the last 0-2 Gyr, the second is 2-4 Gyr, the third is 4-8 Gyr and the bottom is for stars that left clusters more than 8 Gyr ago. The black dots are the locations of the nominal cluster centers. The colors of a cluster's star particles remain the same from panel to panel. 
}
\label{fig_Lage}
\end{figure}

The angular momentum of star stream particles is an approximately conserved quantity, useful for analyzing stream evolution.  The angular momentum of an individual particle oscillates around its orbit in the aspherical potential.    Figure~\ref{fig_Lage} shows the stream particle angular momenta, that is the magnitude of the angular momenta vectors, as a function of the sky longitude for the same stars and age bins as Figure~\ref{fig_hammer}. The blurred structures at large angular momenta are streams orbiting within sub-halos (dwarf galaxies) that are themselves orbiting within the main halo. The near vertical streams at low angular momenta are being tidally disrupted in the inner galaxy.

Streams at low angular momenta are usually closer to the halos center, with a higher sub-halo density, so blur much quicker than those at large angular momenta. Consequently, the thin parts of inner streams are generally younger than those at large radii. Stars of a range of ages in a stream overlap in azimuth, as is most visible for  streams at larger angular momentum. As stars leave a cluster to join a stream with a range of velocities, hence have a range in angular momenta, meaning that at any given angular distance from the progenitor, there is a range in star ages, with the mean age of star particles increasing with distance from the cluster.

\begin{figure}
\begin{center}
\includegraphics[angle=0,scale=0.22,trim=100 20 120 50, clip=true]{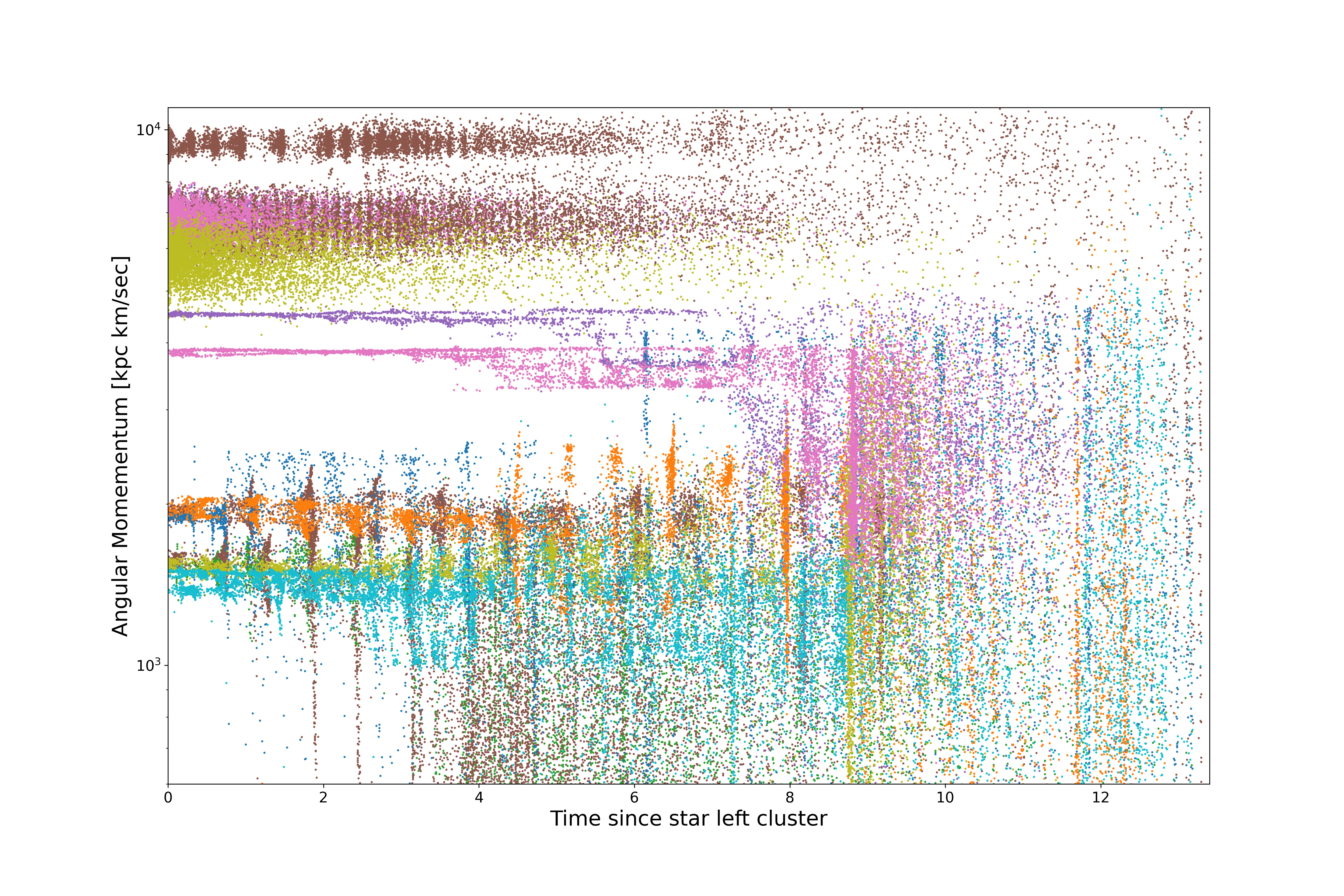}

\vspace{-10pt}
\includegraphics[angle=0,scale=0.22,trim=100 20 120 50, clip=true]{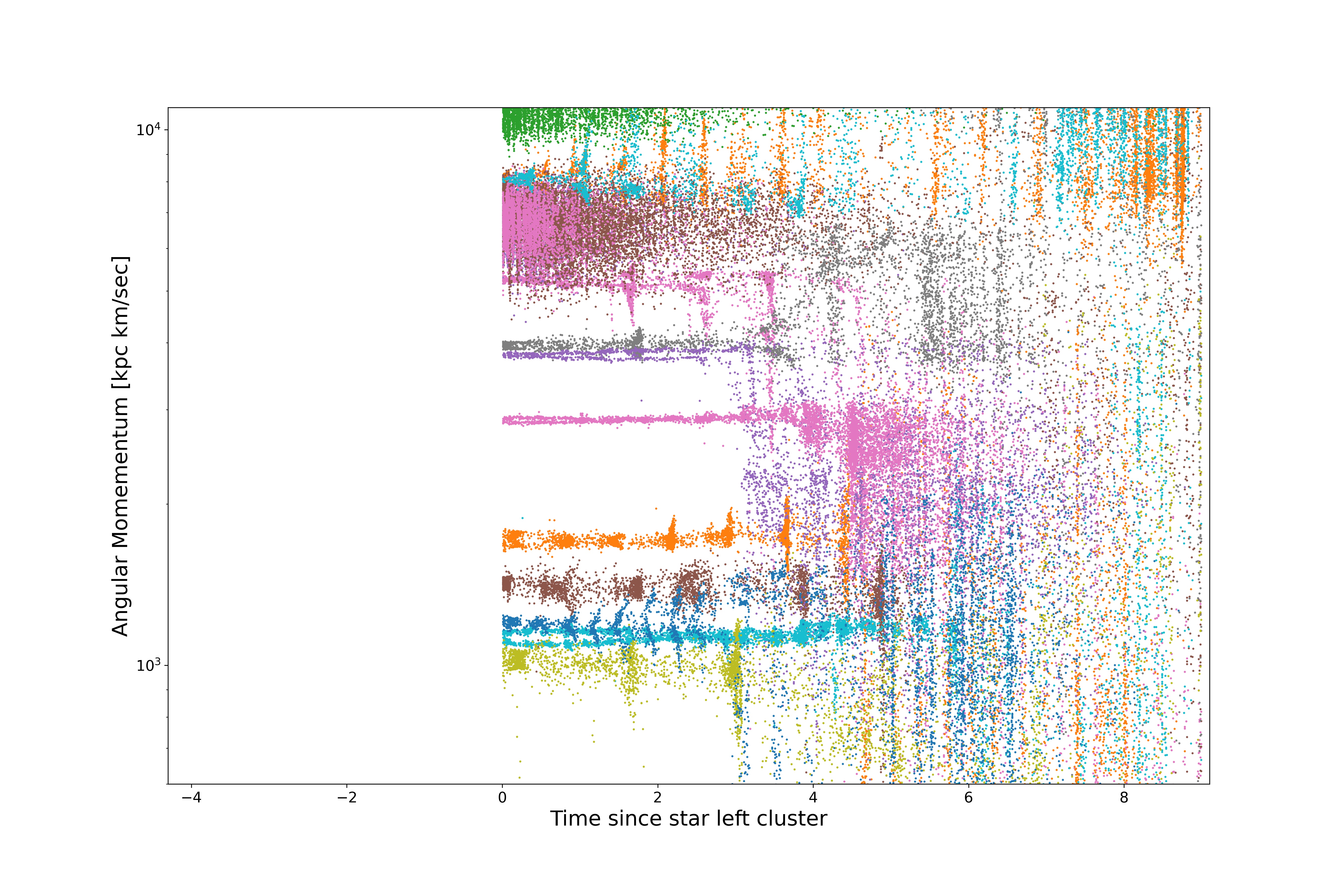}
\end{center}
\caption{The  star particles of Fig.~\ref{fig_Lage} with their angular momenta against the time since the star particle left the cluster. The top panel is for all stars having left the clusters at the end of the simulation; the bottom panel is at time 9.8 Gyr to demonstrate that the angular momentum spreading was similar at earlier times. The clusters have the same colors in both plots, with every third cluster plotted to reduce crowding.
}
\label{fig_Ltime}
\end{figure}

Figure~\ref{fig_Ltime} shows the angular momenta of stream particles versus their stream age for those with clusters in the 10 to 60 kpc radial range. The tidal force broadens the angular momentum distribution of a stream at every pericenter passage, which leads to a spurt of new particles added to a stream  spread in angular momentum. The  low angular momentum clusters have shorter radial orbital periods than higher angular momentum clusters which sets the interval between spurts of particle loss. The range of angular momenta relative to the cluster means that stream particles  that newly join the stream close together in time will spread to a range of angular distances along the stream. 

The panels of Figure~\ref{fig_Ltime} show the angular momentum history for particles with the amount of time they have been in streams. The top panel shows the angular momentum for stars leaving the clusters over the course of the entire simulation and the bottom panel measures the angular momentum at time 9.8 with the 4.3 Gyr younger ages at that time. The figure demonstrates that stars leaving the clusters at earlier times have a similar angular momentum spreading  in time as those at later times. The earliest 3 Gyr show a large spread resulting from the clusters orbiting within their early time sub-halos.

\begin{figure}
\begin{center}
\includegraphics[angle=0,scale=0.55,trim=10 80 40 80, clip=true]{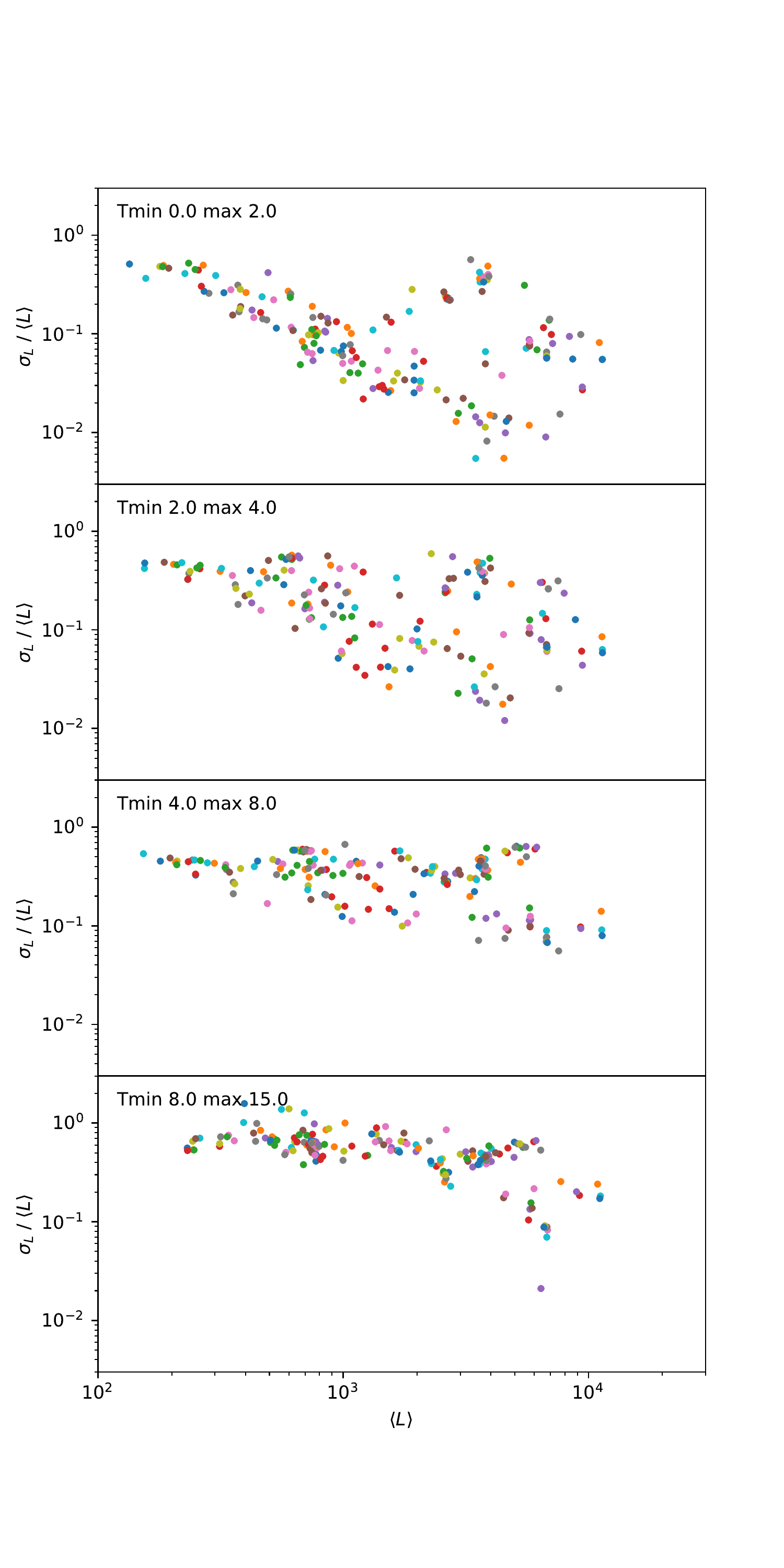}
\end{center}
\caption{The RMS spread in angular momenta of streams, normalized to the mean value for stream particles for particles that left their clusters in the last 2 Gyr (top panel), then 2-4 Gyr, 4-8 Gyr and more than 8 Gyr (bottom panel).
}
\label{fig_fdL}
\end{figure}

Figure~\ref{fig_fdL} shows how the RMS spread of angular momentum steam particles evolves depends on the angular momentum of the stream. The streams at low angular momentum spread quickly with the streams at higher angular momentum, $\sim 10^4$ kpc-\kms\ and above, requiring the entire age of the simulation to be substantially dispersed. Streams at lower angular momentum orbit at smaller radii where the higher sub-halo density leads to higher encounter rates.

\section{Sub-halos and Stream Spreading \label{sec_subhalos}}

\begin{figure}
\begin{center}
\includegraphics[angle=0,scale=0.25,trim=70 40 20 80, clip=true]{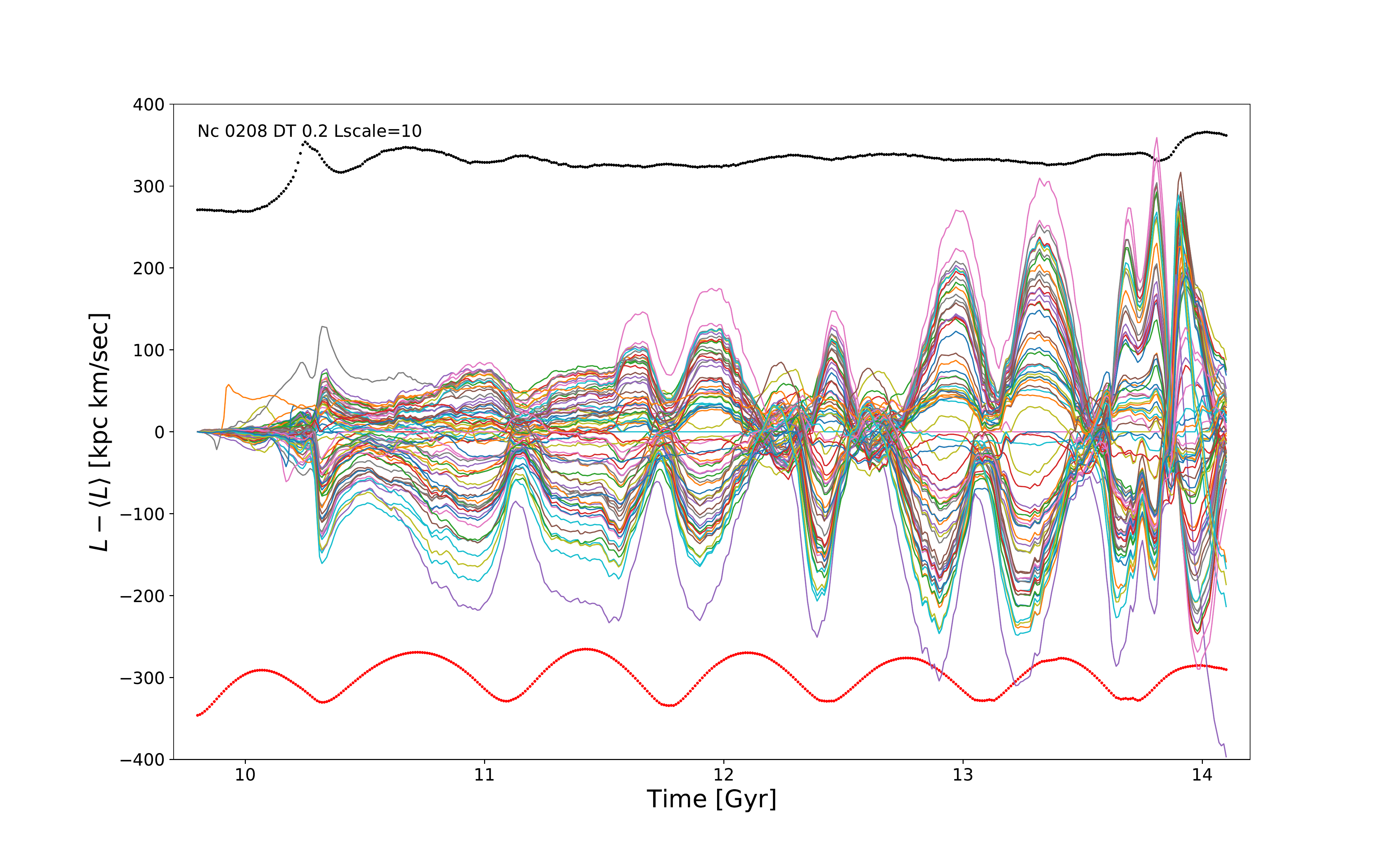}
\includegraphics[angle=0,scale=0.25,trim=70 40 20 80, clip=true]{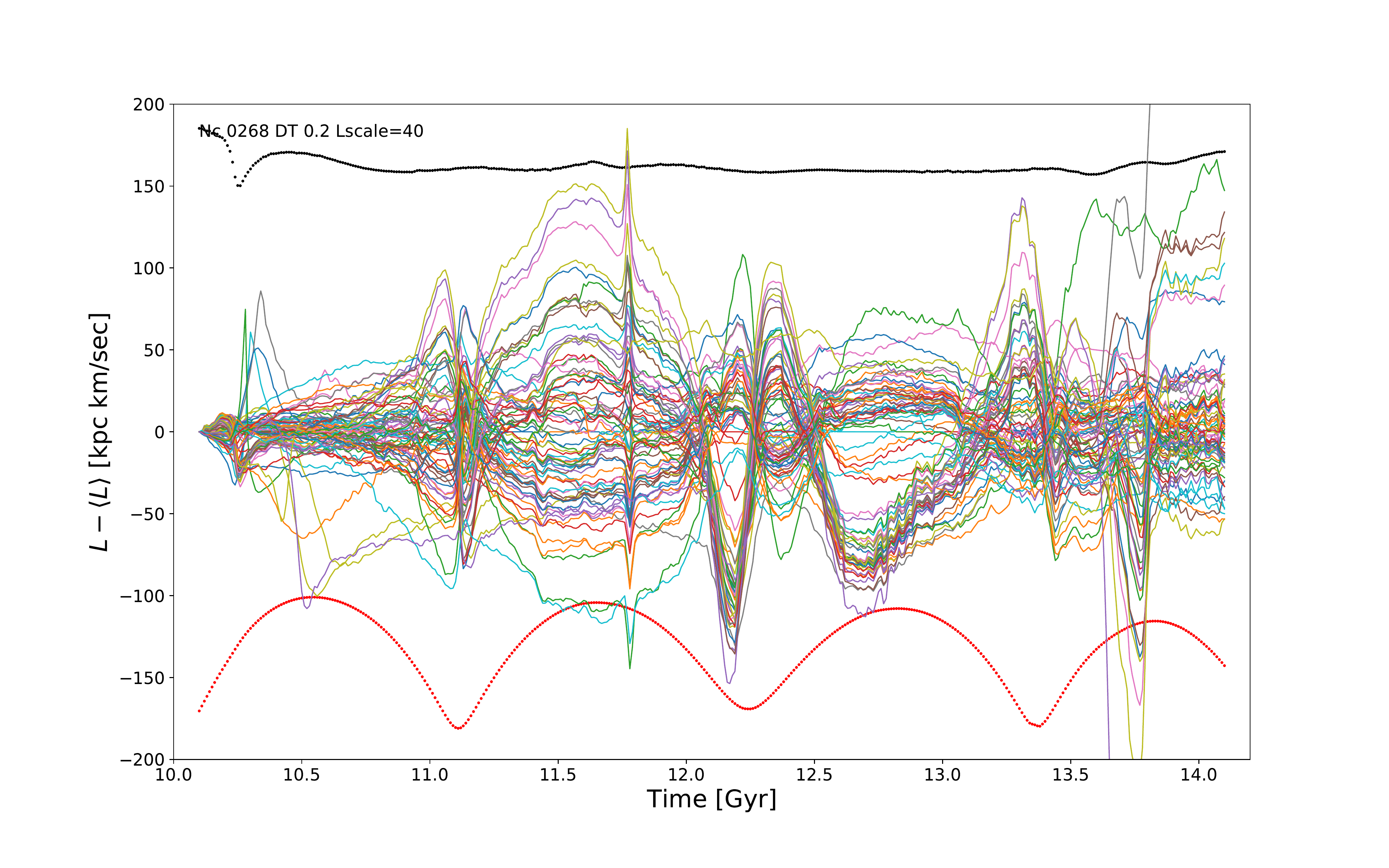}
\end{center}
\caption{The angular momenta for the set of particles released from two clusters between times 9.6 and 9.8 (top) and 9.7 and 10.1 (bottom), both near orbital pericenter of the progenitor cluster. The mean angular momenta (black line) of these two sets of particles are scaled down by 10 and 40, respectively, so have values of approximately 3500 and 6500. The orbits traverse the radial range of approximately 17-33 kpc (top, scaled as $4r-400$)  and 12-60 kpc (bottom, scaled as $1.5r-200$) shown as red dots. 
}
\label{fig_DLT}
\end{figure}

\begin{figure}
\begin{center}
\includegraphics[angle=0,scale=0.37,trim=10 120 60 140, clip=true]{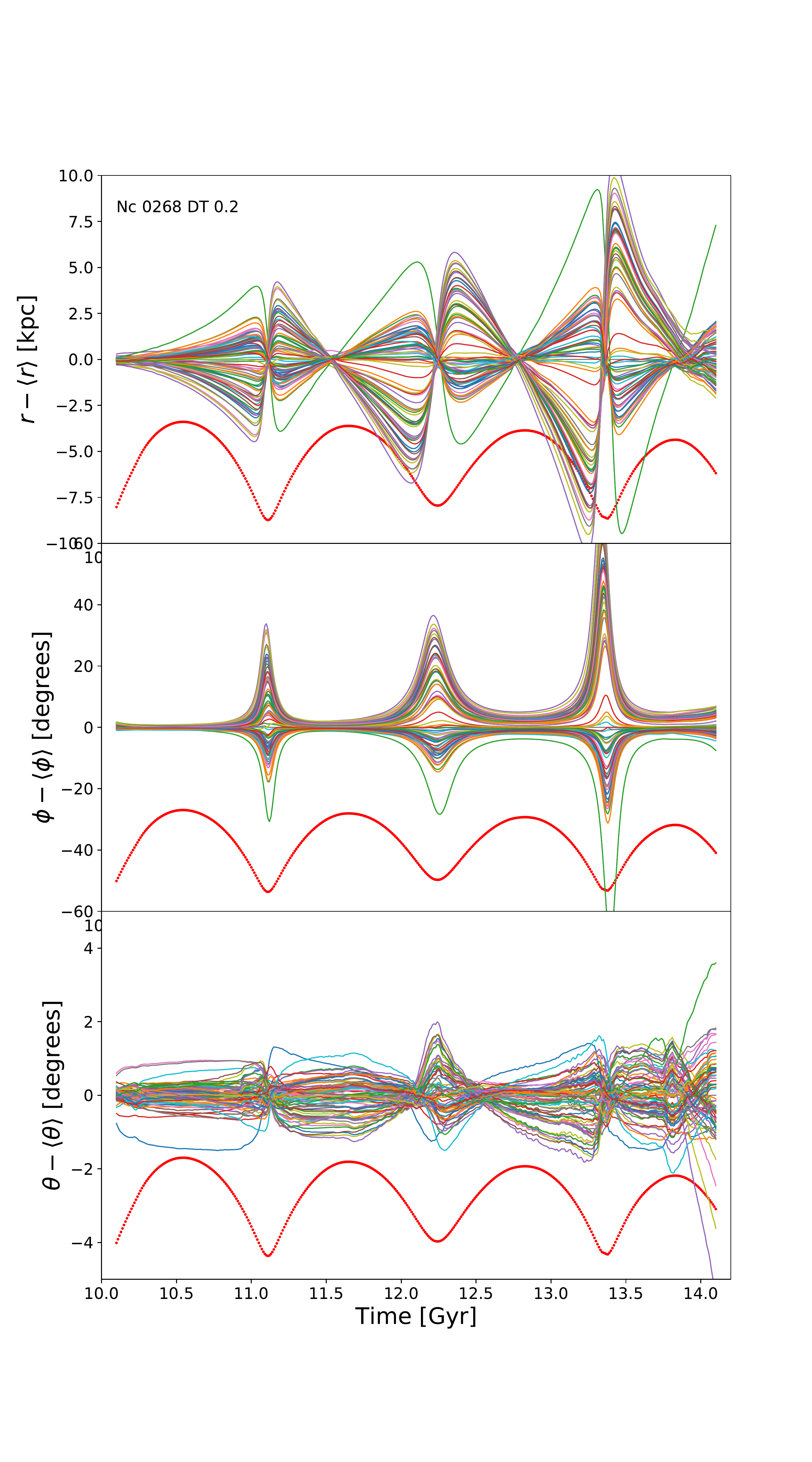}
\end{center}
\caption{The spread in orbital radii (top), azimuth (middle) and latitude (bottom) for the same particles of cluster 268 over the same time interval as in the bottom panel of  Figure~\ref{fig_DLT}. The red lines show the mean orbital radii of the stream segment, scaled as $0.1r-10$, $0.5r-60$ and $0.05r-5$, top to bottom, respectively.
}
\label{fig_DOrb}
\end{figure}

\begin{figure}
\begin{center}
\includegraphics[angle=0,scale=0.25,trim=80 40 100 80, clip=true]{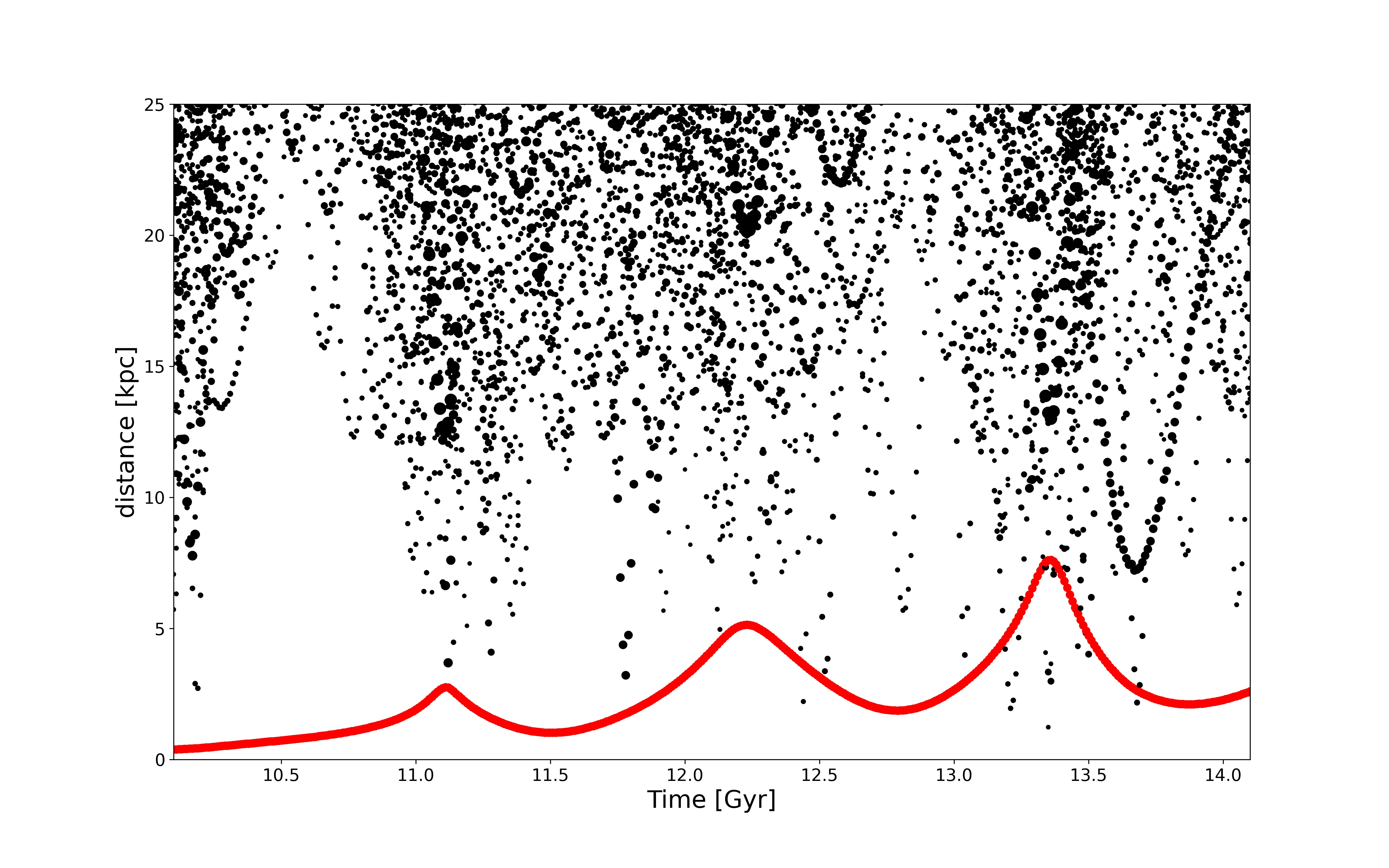}
\end{center}
\caption{The distances between the mean location of the stream segment of the lower panel of Figure~\ref{fig_DLT} and sub-halos more massive than $1\times 10^7 M_\sun$. The dot sizes are proportional to the mass of the sub-halos with the largest dots for the entire mass of the main halo. The red dots are the RMS size of the stream segment. The close encounter at 11.78 Gyr is with a sub-halo of mass $3.6\times 10^8 M_\sun$ moving at a relative velocity of 326 \kms.
}
\label{fig_DHalo}
\end{figure}

\begin{figure}
\begin{center}
\includegraphics[angle=0,scale=0.6,trim=0 60 20 100, clip=true]{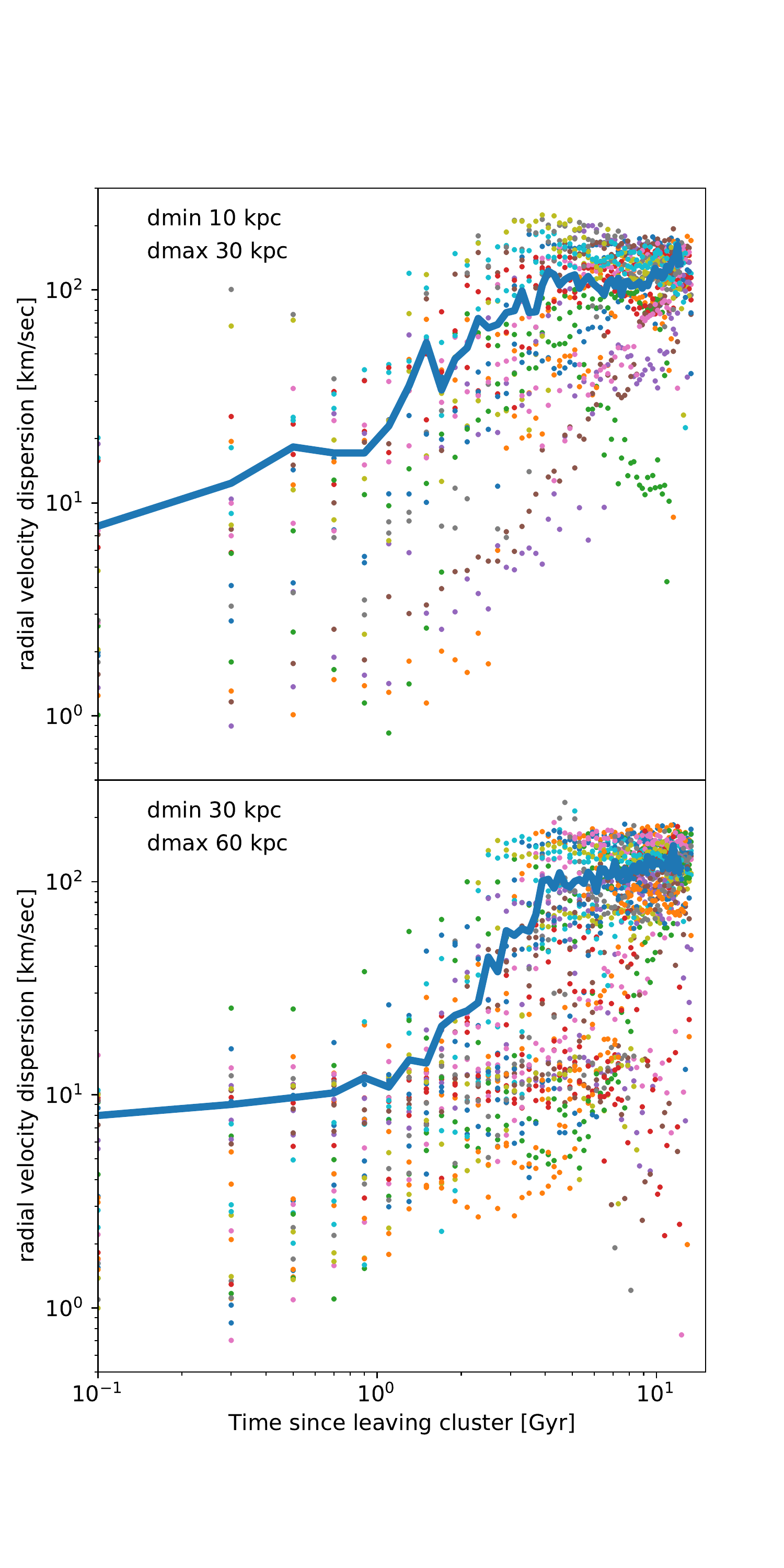}
\end{center}
\caption{The radial velocity dispersion with length of time that the particles have been in the stream, for a set of inner (top) and outer (bottom) streams. The dots have the same color for a given stream, cycling through the streams. The line is the average velocity dispersion at each time bin of 0.2 Gyr wide.
}
\label{fig_sigvrt}
\end{figure}

The particle-by-particle history of the spreading of the angular momenta of stream stars is shown in Figure~\ref{fig_DLT} for two sets of particles emerging over short intervals of time, 0.2 and 0.4 Gyr,  from two clusters (numbers 208 and 228 in Figure~\ref{fig_rotstreams}).  The orbits traverse the radial range of approximately 17-33 kpc (cluster 208)  and 12-60 kpc (cluster 228). Sub-halos pass near and through streams at speeds of 100-200 \kms\ inducing nearly instantaneous velocity changes in the particles of the stream. The fractional changes in angular momentum are typically only 2-3\% in a single encounter, with particles ahead of the point of closest encounter losing angular momentum and those behind gaining. After several orbits the small angular momentum differences develop into significant positional changes, shown in Figure~\ref{fig_DOrb} for cluster 268. The orbits spread a factor of two in radius and azimuth in about 2.5 Gyr, suggesting that the perturbed orbits may have become chaotic and exponentially diverging. Both sub-halo perturbations and orbital spreading in the aspherical, time varying halo potential play a role in spreading thin streams \citep{Ngan16}. The small coherent oscillations of the angular momenta are the result of the motion of the bulge particle at the center of the halo which is used to define the center of momentum of the system.

The mass of the sub-halo that induces the angular momentum spike at time 11.78 in the lower panel of Figure~\ref{fig_DLT} is readily  estimated from the velocity change. The stream segment is at 50 kpc, so the peak velocity change is about 3 \kms. Equation~\ref{eq_deltav} requires a sub-halo of mass $5\times 10^8 (v/300)^{2.04} M_\sun$.  Figure~\ref{fig_DHalo} shows the distances of sub-halos near the stream segment with time. The  sub-halo closest to the stream segment is at 3.2 kpc, moving at  326 \kms, for which Equation~\ref{eq_deltav} gives $3.8\times 10^8 M_\sun$, whereas the directly measured mass of the sub-halo found in the simulation is $3.6 \times 10^8 M_\sun$. with a peak circular velocity of 14.9 \kms\ at 1.7 kpc in the sub-halo.  The rate of sub-halo encounters rises at pericenter, a consequence of the increased angular length of the stream, Figure~\ref{fig_DOrb}. The small wiggles in the angular momenta are the result of the motion of the bulge particle that defines the center of mass.

Figure~\ref{fig_sigvrt} shows the rise in velocity dispersion with time along streams in the 10-30 kpc range and the 30-60 kpc range. Stream particles are binned in ages of 0.2 Gyr with the velocity dispersion of individual streams calculated from the mean subtracted variance. The line is the unweighted average of the stream velocity dispersions. There are an average of about 100 particles per time bin per stream, with a large spread in the numbers. Streams at smaller orbital radii have a quicker  rise in velocity dispersion with time because they are interacting with a higher density of sub-halos, as expected from the calculations of stream heating, and shown in Figure~\ref{fig_heatnum}.   The age of a star in the stream is not an observational quantity, although there is a predicted variation in the stellar mass function with stream age \citep{WebbBovy22}. The distance of a star along the stream is a proxy for stream age, but as shown above, at any given distance there are stars with a wide range of stream ages which blurs the distribution.

\section{Observational tests of sub-halo stream spreading \label{sec_obs}}

Measurement of stream spreading is a potentially powerful test whether the numbers of sub-halos predicted for in a cold dark matter cosmology are present in the Milky Way.  Viable observational tests of stream spreading require the use of practical observational measurements with no knowledge of how long a star has been in the stream. A straightforward approach to a practical measurement is to measure the velocity dispersion with angular distance along the stream from the progenitor.  Another approach is to sum the velocity distribution along the entire length of the visible stream, which does not require any knowledge of the progenitor cluster's location. We will concentrate on the range of 10-60 kpc where currently known streams have been found \citep{Mateu22} and observational velocity measurements with an accuracy of a 1 \kms\ are practical. Spectroscopic observations are important to measure the chemical abundance patterns characteristic of stars from the same globular cluster because streams spread to large angles from the orbital equator of the stream. For potentially practical measurements we use stream stars within 1 or 5\degr\ of the stream equator for our illustrative measurements. 

\begin{figure*}
\begin{center}
\includegraphics[angle=0,scale=0.43,trim=110 70 20 40, clip=true]{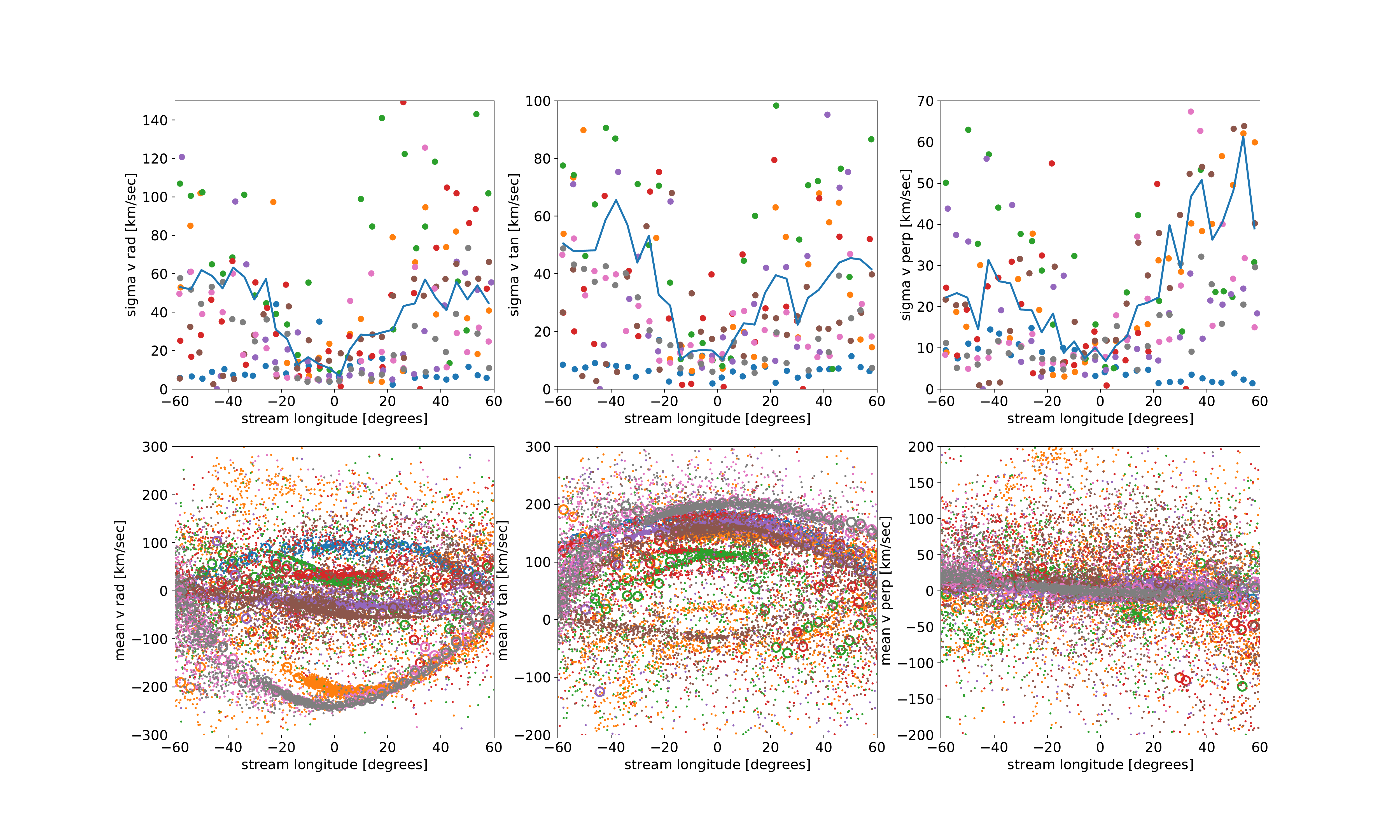}
\end{center}
\caption{The rise in velocity dispersion with angular distance along the stream from the progenitor cluster. The top row of panels shows the three components of the velocity dispersion for stars within 5\degr\ of the cluster orbit centerline for clusters in the range 10-30 kpc, with $L>2000$ kpc-\kms. The data are binned in stream longitude bins of 4\degr. The lines in the upper panels are the averages of the velocity dispersions. The bottom row shows the star particle velocities as points and the mean velocities along a stream as open circles.  Velocities are measured in the great circle frame of the progenitor cluster, without removing its velocity.
\label{fig_vsigvd}}
\end{figure*}

\subsection{Velocities with Stream Longitude}

The distribution of velocity dispersions of stream stars  is shown as a function of angular distance for the leading (positive angles) and trailing tidal arms for the 8 streams in the baseline simulation within the 10-30 kpc with $L>2000$ kpc-\kms\ are shown in Figure~\ref{fig_vsigvd}. Lower angular momenta streams come close to the disk which provides a large scale tide that further disperses the streams.  The velocities are measured in spherical coordinates in each stream's great circle frame, where the cluster angular momentum defines the orbital plane and its position the zero of stream longitude and latitude. The plots show stars within 5\degr\ of the orbital equator. Using stars only within 1\degr\ of the stream equator suppresses the rise in velocity dispersion beyond angles 10-20\degr\ along the stream. Including streams stars out to a width 20\degr\ gives a $\simeq$50\% increase in the velocity dispersion in all components.   The mean and standard deviations of the velocities are made in bins of 4\degr\ along the streams.   Measurements in half the width and half the length give very similar results, but with doubled errors. In these measurements all orbital phases of streams are taken and there is no tracing of the high density centerline of the stream with length. 

Figure~\ref{fig_vsigvd} shows that the radial velocity dispersion is generally larger than tangential  or perpendicular velocity dispersion. The tidal forces that pull stars away from the clusters are essentially radial, giving the stars a larger initial radial velocity dispersion than in the tangential or perpendicular directions. The primary heating action of sub-halos is to pull stars on either side of a crossing point toward the sub-halo, which causes the tangential and vertical motions to have a more straightforward dependence on sub-halo stream heating.  The velocity dispersions in the leading and trailing streams are sufficiently symmetric that they are folded together to increase the statistical significance of simple $\sigma \propto \phi^\alpha$ fit with angular distance, $\phi$. 

\begin{table}
\caption{Sub-halo and Stream Counts \label{tab_ss}}
\begin{tabular} {| r | r |r|}
\tableline
Model & $N_{\rm halos} $ &$N_{\rm streams}$  \\
\tableline
  disk & \multicolumn{2}{|c|}{10-60 kpc}  \\
\tableline
baseline & $39\pm4$ & 29\\
no  disk & $54\pm4$ & 66  \\
2$\times$ disk & $31\pm3$ & 36 \\
\tableline
\end{tabular}
\end{table}

The  particle velocities along the streams are shown in the lower row of panels of Figure~\ref{fig_vsigvd}. The mean radial and tangential velocity along the stream, shown as the open circles, primarily reflect the orbital motion along the various streams. The mean vertical velocity along the stream (the latitude component of the velocity in the stream's great circle frame)  is an indicator that streams can be tilted with respect to the orbital equator and that there can be relatively large scale velocity perturbations from the more massive orbiting sub-halos \citep{Erkal19,Shipp19}. The velocity dispersions along the streams shown in Figure~\ref{fig_vsigvd} has a lot of  scatter, as should be expected for the essentially random process of sub-halo encounters with streams. For the 29 streams identified in the 10-60 kpc region the average radial velocity dispersion rises  from 10 \kms\ to 50 \kms\ in 20-30\degr, the tangential velocity dispersion rises from 7 \kms\ to 40 \kms\ in 40\degr, and the vertical velocity dispersion from 7\kms\ to 25\kms\ in 40\degr.  

\begin{figure*}
\begin{center}
\includegraphics[angle=0,scale=0.42,trim=110 10 20 40, clip=true]{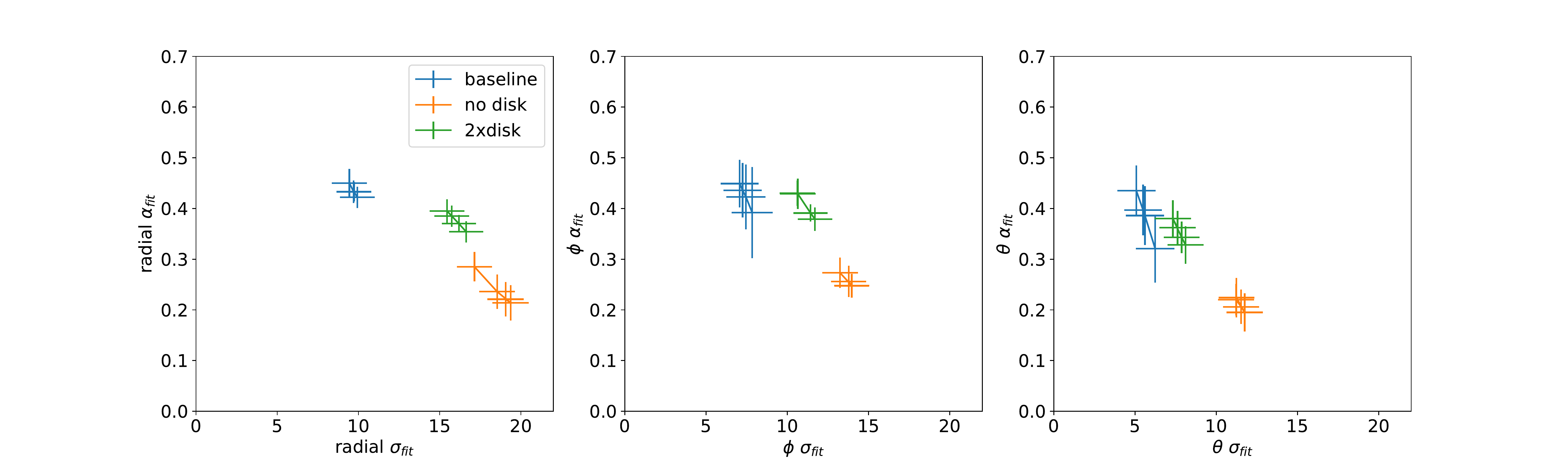}
\end{center}
\caption{The fits to the $\sigma = \sigma_{fit} \phi^\alpha$ relation for the three components of the velocity, with  the 68\% probability error bars.  Fits extending  longitudes of 30-60\degr\ are shown for each set of streams.  All fits are done for velocities measured within $\pm$5\degr\ of the stream center.
}
\label{fig_sigfits}
\end{figure*}

\begin{figure*}
\begin{center}
\includegraphics[angle=0,scale=0.43,trim=140 0 120 20, clip=true]{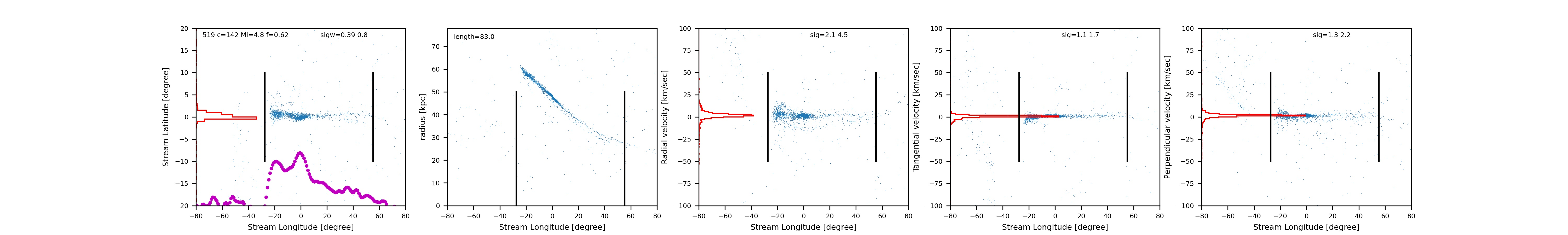}
\includegraphics[angle=0,scale=0.43,trim=140 0 120 20, clip=true]{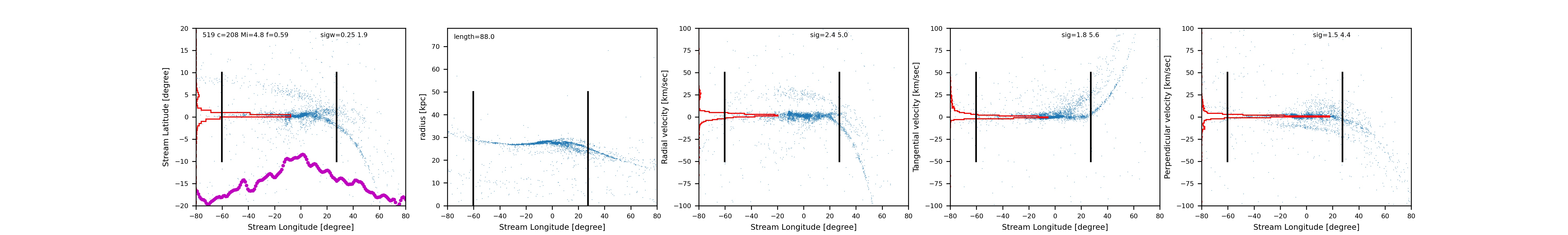}
\includegraphics[angle=0,scale=0.43,trim=140 0 120 20, clip=true]{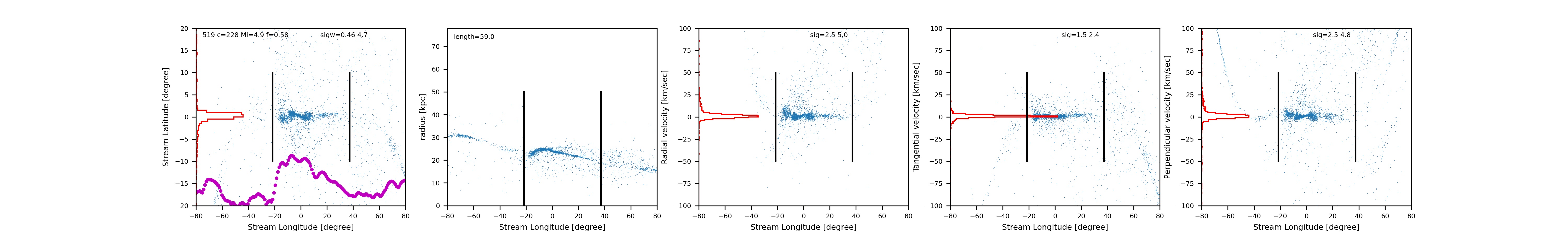}
\end{center}
\caption{The streams in the 10-60 kpc range longer than 50\degr\ after fitting  with a fourth order polynomial to remove large scale path variations. From left to right the panels show the density on the sky, the galactocentric radii, the radial, tangential, and perpendicular velocities. The histogram on the vertical axis gives the distribution of the particles within the measured length of the stream, as indicated by the vertical bars. The magenta points in the first panel give the logarithm of the peak mass surface density, $D_p$, along the stream, scaled as $5\log_{10}({D_p})-20$. A typical maximum peak density is $300 M_\sun /\square\degr$ . Stream ends  are marked with vertical black lines. The 3$\sigma$ and 6$\sigma$ clipped velocity dispersions along the stream are at the top of the 3 velocity panels.
}
\label{fig_ssfit}
\end{figure*}

Table~\ref{tab_ss} reports the  number of sub-halos of mass $10^{7-8} M_\sun$  in the 10-60 kpc range in the second column and the number of streams in the same radial range in the third column.   The double mass disk has a few more sub-halos than the baseline model at the final moment measured, although the time average over the last 1.5 Gyr is lower than the baseline.

Figure~\ref{fig_sigfits} plots the coefficients of the power law to  the, $\sigma-\phi$ velocity dispersion longitude relation for the three components of the velocity.  The fits are done as least squares to the linear logarithmic relation, $\log{\sigma_{fit}} + \alpha_{fit} \phi$.  The plotted results show a weak dependence on the maximum longitude fitted, which  ranges from 30-60\degr. Each simulation has a distinct set of $\sigma-\phi$ relations. The simulation without a disk has the most sub-halos and has the largest velocity dispersion. The baseline and the double disk simulation have similar $\phi$ dependence with the double disk model, which has more sub-halos at late times, Table~\ref{tab_ss}, having larger velocity dispersions.  The  measurements of the velocity dispersions of a population of streams offer some prospect for estimating the numbers of sub-halos within a galactic halo. The results are sensitive to which streams are selected and how the measurements are made so will need comparison to a set of simulations. The most challenging aspect for these measurements is that the stream progenitor needs to be identified, which is not  possible for many streams \citep{Bonaca21}.

\begin{figure*}
\begin{center}
\includegraphics[angle=0,scale=0.35,trim=190 10 160 20, clip=true]{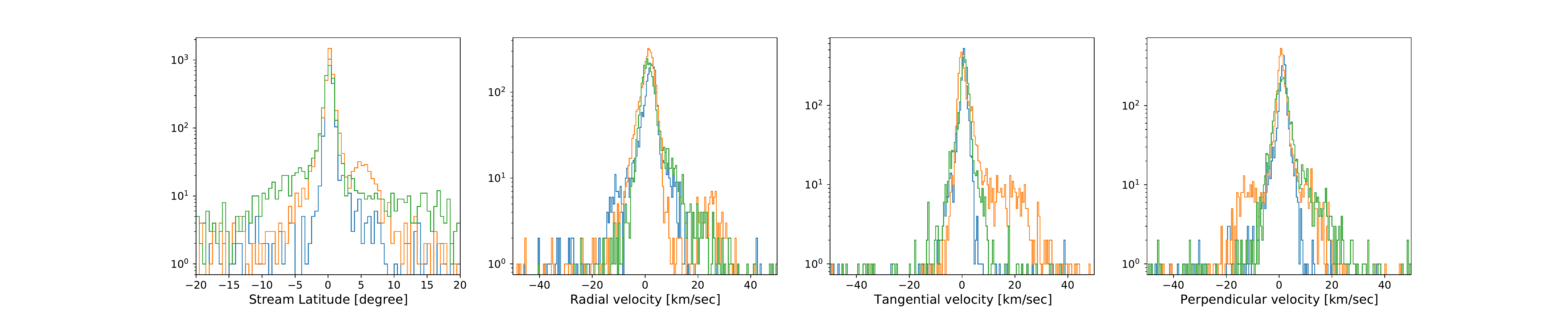}
\includegraphics[angle=0,scale=0.35,trim=190 10 160 20, clip=true]{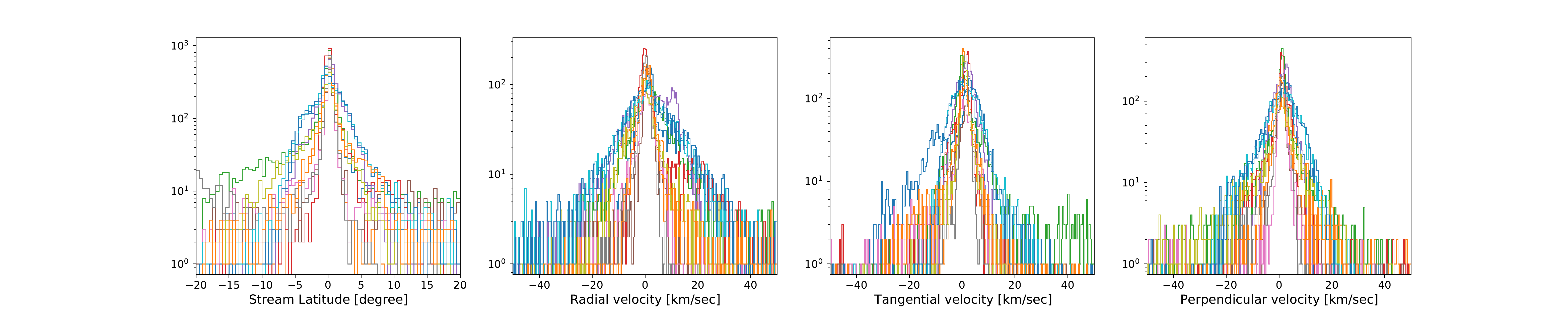}
\end{center}
\caption{The logarithmic distribution of the particles within $\pm$5\degr\ of the stream centerlines in latitude, radial velocity, tangential velocity and perpendicular velocity, in the  left to right panels, respectively, for streams longer than 50\degr. The top  panels are for the baseline and bottom panels are for the no-disk simulations.
}
\label{fig_hist}
\end{figure*}

\subsection{Longitude Summed Velocity Distributions}

The distribution of velocities along a set of streams does not require any knowledge of the location of the progenitor, summing the velocities of stars of all  longitudes along the visible length of a stream. Figure~\ref{fig_ssfit} shows a set of streams longer than 50\degr\ in the 10-60 kpc range.  These simulated streams are approximate analogs for the well-studied GD-1 stream \citep{GD1}.  The stream particles are placed into a grid of cells ($1\degr \times 0.5\degr$ on the sky) and smoothed with a 2D Gaussian, $\sigma_\phi=1.0\degr$, $\sigma_\theta=0.75\degr$. Streams are required to have a standard deviation of the longitudes of their particles of more than 20 \degr, which removes the streams with many stars in a fuzzy distribution around the progenitor. The cells above a threshold of 2 $M_\sun / \square\degr$ are fitted to a fourth order polynomial.  The velocities in each component are fitted with the same procedure. The panels in each row show the particle latitude offsets from the mean stream, radius (no streamline correction) , the offsets in radial velocity, tangential velocity, and perpendicular velocities with stream longitude on the sky. In the leftmost panel the logarithm of the filtered peak density along the centerline is inserted into the figure with a scaling to the axis values of $5\log_{10}(D_p)-15$.  

 Figure~\ref{fig_hist} shows  the star particle distributions in streamline fit-adjusted latitude, radial velocity, tangential velocity, and velocity in the latitude direction, from left to right, respectively, for the three streams longer than 50\degr\ plotted in Figure~\ref{fig_ssfit}.  Streams having an RMS spread in longitude greater than 10\degr\ are analyzed. Stream ends are defined as where the surface density drops below  $2 M_\sun /\square\degr$ or the stream path changes more than 2\degr\ or the velocity changes more than $30 \kms$. The entire population of 12 streams longer than 10\degr\ has similar values. The velocity distributions all have comparably narrow cores of $\sim 2-5$ \kms, with broad non-Gaussian wings that vary substantially from stream to stream.   The longer than 50\degr\  streams in the  baseline simulation have a mean radial velocity dispersion of  2.1 \kms\ within $\pm$1\degr  GD-1 has a FWHM width of 0.5\degr \citep{Koposov10}, that is a 1$\sigma$ width of 0.21\degr, and a measured radial velocity dispersion of  2.1$\pm$0.3 \kms\  \citep{GD1sigma21}. The simulation without a disk has long streams that are more than twice the width and velocity dispersion of the baseline simulation. The simulations convincingly show that the greater sub-halo numbers in a simulation without a galactic disk leads to significantly greater sub-halo heating of the streams. The baseline simulation streams have widths and velocities comparable to GD-1, but a statistical test requires consideration of observational selection effects and should be done with more simulations.

\begin{figure*}
\begin{center}
\includegraphics[angle=0,scale=0.42,trim=130 0 0 0, clip=true]{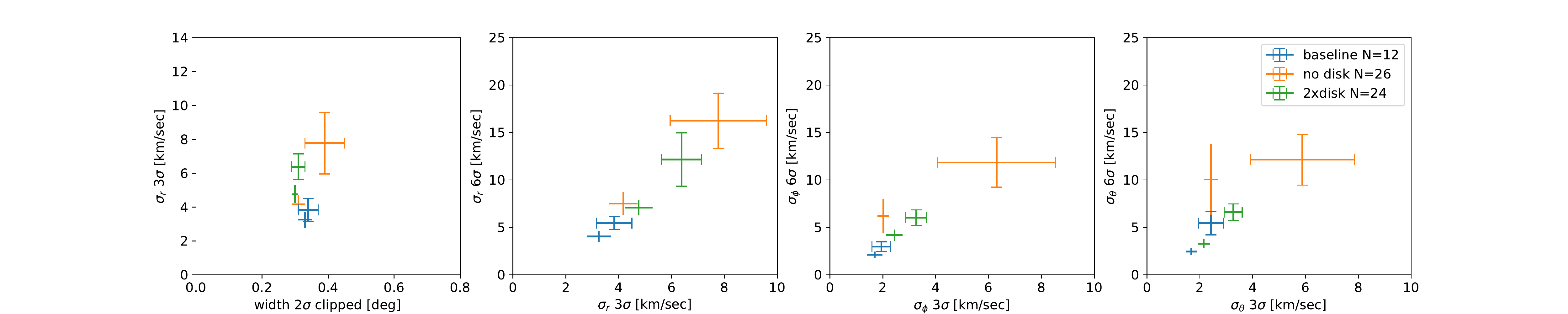}
\includegraphics[angle=0,scale=0.42,trim=130 0 0 0, clip=true]{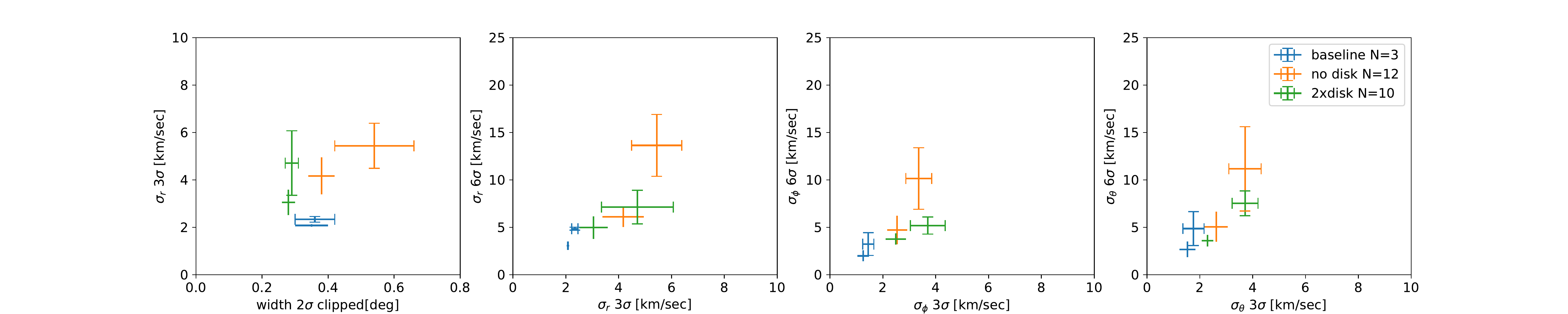}
\end{center}
\caption{The $2\sigma$ widths, 3$\sigma$, and 6$\sigma$ mean velocity dispersions along all streams longer than 10\degr\ (top row) and 50\degr\ (bottom row) in the three simulations. The error bars are the standard deviation of the mean. The measurements are made within 1\degr\ and  5\degr\ of the streamline, the latter having caps on the error bars. The measurement does not require a progenitor location.
}
\label{fig_sig36}
\end{figure*}

The stream fitting procedure used for the long streams of Figure~\ref{fig_ssfit} is applied to all streams longer than 10\degr. The 3$\sigma$  and 6$\sigma$ velocity dispersions are measured for each stream. The average 3$\sigma$  and 6$\sigma$ clipped velocity dispersions are displayed in Figure~\ref{fig_sig36} with the error bars showing the standard deviation of the mean values. The number of streams in the average is given in the legend of the plot. The differences between the streams are significant in the average over the stream population. An important caveat is that these streams all had progenitors with masses above $4\times 10^4 M_\sun$, with a typical cluster losing about 60\% of its mass. Therefore yet lower mass clusters, which lose more mass earlier, are likely to have even higher velocity dispersions. 

Velocity dispersions measured within 1\degr\ of the stream are most comparable to current velocity measurements, which find a radial velocity dispersion of $2.1\pm 0.3$ \kms\  \citep{GD1sigma21}. The average radial velocity dispersions of streams longer than 50\degr\ in the simulations are 2.1, 3.0, and 4.2 \kms\ for the models with our baseline disk, the double mass disk, and the no-disk model, respectively.  All three of the baseline disk simulation have velocity dispersions below 2.55 \kms, whereas the no-disk model 4 of the 12 long streams are below this limit. A single stream is indicative of the sub-halo population, but confident conclusions require study of the entire population of streams.

\section{Discussion and Conclusions\label{sec_discussion}}

The dark matter sub-halos orbiting within the galactic halo perturb the velocities of stars along the streams, which increases their velocity dispersion.  The result is a dramatic increase in the spread of stream stars on the sky with time since they left their progenitor cluster. Sub-halos in approximately the $ 10^{7-8} M_\sun$ range dominate stream heating, based on their numbers and an impact approximation scattering calculation. There is considerable stream to stream variation because there are typically only 30-50 sub-halos of the dynamically important $10^{7-8} M_\sun$ mass interval in the 10-60 kpc region where streams are currently found. The divergence of orbits in the aspherical potential adds to the stream spreading. 

Our simulations with disks contain streams with FWHM and velocity dispersions comparable to those observed, although a statistical comparison to the stream population allowing for observational selection effects and field star rejection procedures has yet to be done.  Most notably there are thin streams longer than 50\degr. For those streams the average 3$\sigma$ clipped radial velocity dispersion within $\pm$1\degr\ is  2.1 \kms\ in the baseline simulation rising to 4.2 \kms\ in the no-disk simulation. The 3$\sigma$  and 6$\sigma$ clipped velocity dispersions have ratios ranging around 2. The significantly non-Gaussian velocity and width distributions are signature indicators of the sub-halo heating of streams over their lifetimes.

The angular distance of stars along the stream from the progenitor cluster is a proxy for the time a star has been in the stream, although there is a spread of ages at any given distance. We measure the velocities within $\pm$5\degr\, because restricting the measurements to stars within  $\pm 1\degr$ of the stream equator misses many higher velocity stars.   We fit the streams in  the 10-60 kpc region above a minimum angular momentum of 2000 kpc-\kms\ to avoid streams that come close to the disk.  Figure~\ref{fig_sigfits} presents fits for the components of the velocity dispersion with distance along the streams. Average velocity dispersions rise from $\simeq$3-5 \kms\ near the clusters to $\simeq$30-40 \kms\ at 20-30\degr\ away. The velocity dispersions depend on the numbers of sub-halos in the same volume as the streams, with $\sim$50\% higher velocity dispersion in a model with no disk which leaves about 60\% more sub-halos in the inner 60 kpc, Table~\ref{tab_ss}.

The distribution of stars in position and velocity  within a strip close to the stream above a minimum density threshold are measurements that do not require locating the progenitor star cluster.  The stream widths and velocity dispersions  plotted in Figure~\ref{fig_sig36} show that the no-disk simulation with the most sub-halos has the highest velocity dispersion streams. The statistical significance of the differences are greater for measurements extending 5\degr\ from the stream. The lower panel of Figure~\ref{fig_sig36}  shows that  the long streams in the baseline simulation have a radial velocity dispersion of 2.0 \kms whereas the disk-less simulation streams have an average radial velocity dispersion of 4.2 \kms,  twice that of measurements of GD-1 \citep{GD1sigma21}. However, 4 of the 12 no-disk long streams do have velocity dispersions comparable to GD-1. That is, a few streams with $\sim2$ \kms\ radial  velocity dispersions are present in all three simulations. These results underscore that unbiased velocity measurements around a set of Milky Way streams will be required to reach confident conclusions about the sub-halo population.

\begin{acknowledgements}
Comments from an anonymous referee substantially improved this paper, as did comments from Ana Bonaca, Kathryn Johnston and David Weinberg.

This research was supported by NSERC of Canada. Computations were performed on the niagara supercomputer at the SciNet HPC Consortium. SciNet is funded by: the Canada Foundation for Innovation; the Government of Ontario; Ontario Research Fund - Research Excellence; and the University of Toronto.
\end{acknowledgements}

\software{Gadget4: \citet{Gadget4}, Amiga Halo Finder: \citep{AHF1,AHF2}}

\bibliography{Stream}{}
\bibliographystyle{aasjournal}

\end{document}